\documentclass[preprint,number]{elsarticle} 
\usepackage{url}
\usepackage[english]{babel}
\usepackage{graphicx}
\usepackage{amsmath}
\usepackage{bbold}
\usepackage{color}
\usepackage{longtable}
\biboptions{sort&compress} 
\usepackage{amsthm}
\usepackage{enumitem}
\usepackage{subcaption}
\usepackage{todonotes}
\usepackage{booktabs,tabularx,array}
\usepackage[skip=4pt]{caption}
\newcolumntype{Y}{>{\raggedright\arraybackslash}X}
\renewcommand{\arraystretch}{1}
\usepackage{xurl}
\theoremstyle{remark}
\newtheorem*{remark}{Remark}
\usepackage{fancyhdr}
\usepackage{etoolbox}
\patchcmd{\maketitle}{\thispagestyle{empty}}{\thispagestyle{fancy}}{}{}

\begin{document}
\begin{frontmatter}

\title{Cross-Atlantic Research Agenda for Scalable Grid Architectures and Distributed Flexibility}

\author[uvm]{Mads R. Almassalkhi\corref{cor1}}
\author[uvm]{Dakota Hamilton}
\author[jhu]{Hasan Giray Oral}
\author[jhu]{Yury Dvorkin}
\author[jhu]{Dennice Gayme}
\author[cu]{Bri-Mathias Hodge}
\author[aau]{Brian Vad Mathiesen}
\author[aau]{Jakob Stoustrup}
\author[dtu]{Tobias Ritschel}
\author[dtu]{Rune G. Junker}
\author[dtu]{Shahab Tohidi}
\author[dtu]{Razgar Ebrahimy}
\author[dtu]{Henrik Madsen\corref{cor2}}
\address[uvm]{University of Vermont, Vermont, USA}
\address[jhu]{Johns Hopkins University, Maryland, USA}
\address[cu]{University of Colorado Boulder, Colorado, USA}
\address[aau]{Aalborg University, Denmark}
\address[dtu]{Technical University of Denmark, Denmark}


\begin{abstract}
{\color{black} 
Electric power systems are rapidly evolving into deeply digital, cyber-physical infrastructures in which large fleets of distributed energy resources must be coordinated as system-level flexibility across multiple spatial and temporal scales. 
Despite growing distributed energy resource deployment, existing grid and market architectures lack scalable, interoperable mechanisms to reliably translate device-level flexibility into grid-aware services, creating risks to reliability, affordability, and resilience at high penetration. 
We propose that scalable and reliable coordination of distributed energy resource-based flexibility in future power systems is fundamentally an architectural problem that can be addressed through laminar cyber-physical design using minimal, standardized interoperability interfaces that link device autonomy with system-level objectives. 
To assess this claim, we present and discuss a layered cyber-physical systems architecture and explicate its implementation through standards-based interfaces, Flexibility Functions, hierarchical control, and case studies spanning U.S. and Danish regulatory, market, and operational contexts. 
Empirical evidence from New York’s Grid of the Future proceedings, Danish Smart Energy Operating System pilots, and operational aggregator deployments demonstrates that such architecture enables predictable, grid-aware flexibility while preserving device autonomy, interoperability, reliability, and quality of service. 
These results support a cross-Atlantic research agenda centered on joint testbeds, harmonized interoperability mechanisms, and coordinated policy experiments to accelerate the deployment of resilient, scalable, and flexible clean energy systems.
}
\end{abstract}

\end{frontmatter}




\clearpage

{\color{black}
\section*{List of Abbreviations and Acronyms}
\begin{longtable}{@{}ll@{}}
ADMS & Advanced Distribution Management System \\
AI & Artificial Intelligence \\
aFRR & Automatic Frequency Restoration Reserve \\
AMI & Advanced Metering Infrastructure \\
API & Application Programming Interface \\
BTM & Behind-the-Meter \\
CAISO & California Independent System Operator \\
C\&I & commercial and industrial \\
DER & Distributed Energy Resource \\
DERMS & Distributed Energy Resource Management System \\
DK1 & Danish bidding zone 1 (label) \\
DG & Distributed Generation \\
DLMP & Distribution Locational Marginal Price \\
DNO & Distribution Network Operator \\
DR & Demand Response \\
DRAM & Demand Response Auction Mechanism \\
DSF & Demand-Side Flexibility \\
DSIP & Distribution System Implementation Plan \\
DSO & Distribution System Operator \\
DSSE & Distribution System State Estimation \\
ERCOT &  Electric Reliability Council of Texas \\
EV  & Electric Vehicle \\
EU & European Union \\
FACTS & Flexible AC Transmission System \\
FCR & Frequency Containment Reserve \\
FERC & Federal Energy Regulatory Commission \\
FF & Flexibility Function \\
FTM & Front-of-the-Meter \\
GET & Grid-Enhancing Technology \\
GOTF & (New York) Grid of the Future  \\
HEMS & Home Energy Management System \\
HMIS & Home Management Information System \\
HVAC & Heating, Ventilation, and Air Conditioning \\
IEC & International Electrotechnical Commission \\
IEEE & Institute of Electrical and Electronics Engineers \\
IoT & Internet-of-Things \\
ISO & Independent System Operator \\
LV/MV/HV & Low-Voltage / Medium-Voltage / High-Voltage \\
mFRR & Manual Frequency Restoration Reserve \\
MIM & Minimum Interoperability Mechanism \\
MISO & Midcontinent ISO \\
MOU & Memorandum of Understanding \\
M\&V & Measurement and Verification \\
NERC & North American Electric Reliability Corporation \\
NIST & National Institute of Standards and Technology \\
NYISO & New York ISO \\
NYSERDA &  New York State Energy Research and Development Authority \\
OASC & Open \& Agile Smart Cities \\
OEM & Original Equipment Manufacturer \\
PEM & Packetized Energy Management \\
PJM & Pennsylvania-New Jersey-Maryland Interconnection \\
PV & Photovoltaics \\
RES & Renewable Energy Source \\
RTO & Regional Transmission Organization \\
RTU & Remote Terminal Unit \\
SCADA & Supervisory Control and Data Acquisition \\
SE-OS & Smart Energy Operating System \\
SOC & State-of-Charge \\
TSO & Transmission System Operator \\
UFLS & Underfrequency Load Shedding \\
VELCO & Vermont Electric Power Company \\
VPP & Virtual Power Plant
\end{longtable}
}

\clearpage

\section{Introduction: Motivation and Scope}

\begin{figure}[t]
    \centering
    \includegraphics[width=0.67\linewidth]{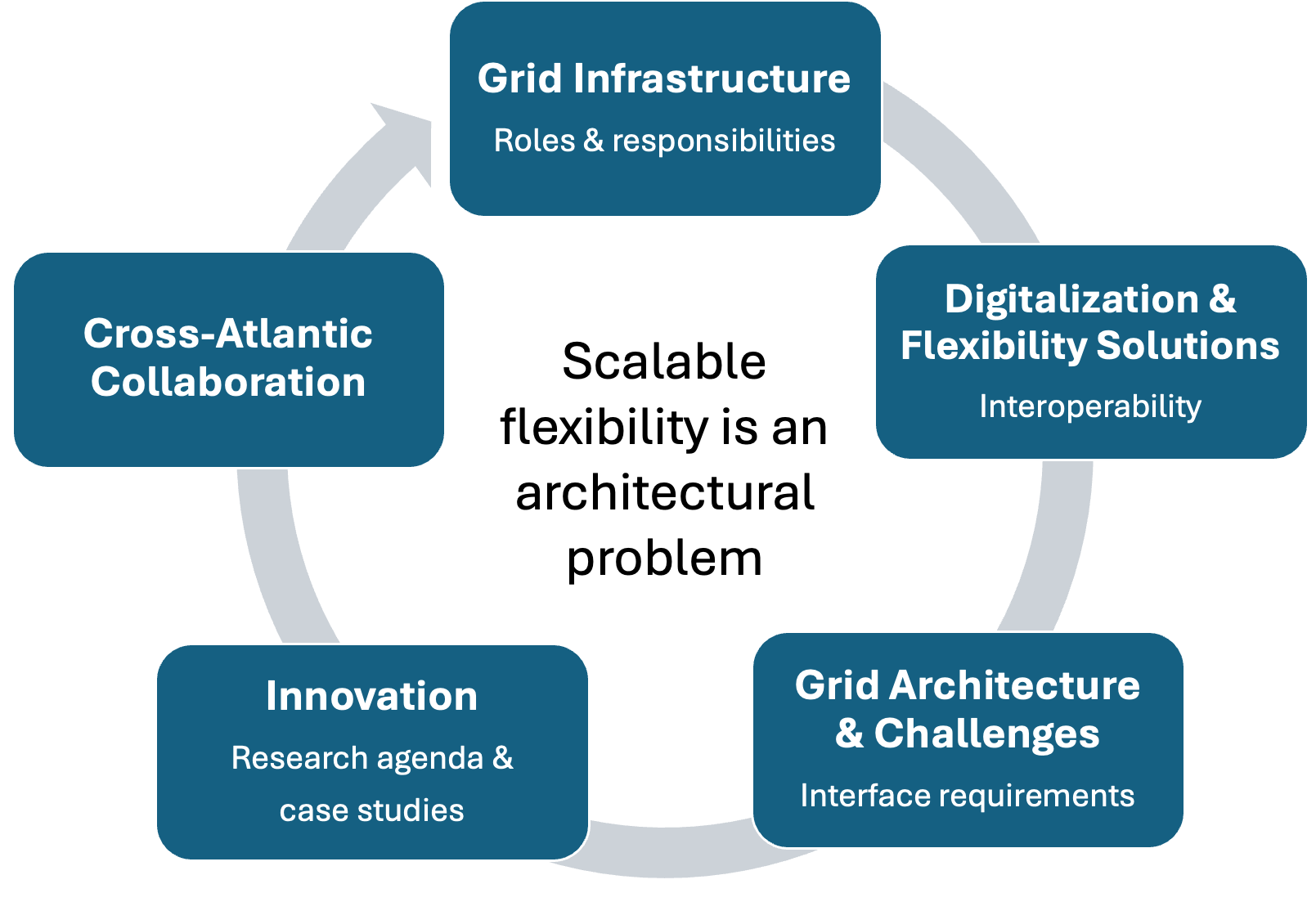}
    \caption{{\color{black}A graphical overview of the paper's themes.}}
    \label{fig:paperoverview}
\end{figure}

Efforts to mitigate and adapt to climate change, the increasing cost-competitiveness of low-emissions generation, electrification of transportation, heating, and cooling, and the need to reduce reliance on hostile or unreliable energy imports have strained electricity supply. At the same time, economy-wide artificial intelligence (AI) advances and the (unexpectedly rapid) emergence of data centers continue to drive electricity demand. Although these factors manifest differently around the globe, they undermine electricity affordability, supply reliability, public support for clean energy and climate goals, and, as a result, the overall \textit{achievability} of these goals.
These goals generally prescribe minimum targets for clean, distributed energy generation and consumption and a reduction in CO$_2$ emissions by a desired target year (e.g., Denmark seeks to reduce CO$_2$ emissions by 70\% by 2030 \cite{denmark2020greensustainableworld}, while Vermont targets 80\% reduction (relative to 1990 baselines) by 2050~\cite{vermont2020gwsact}). Combating climate change and enhancing energy security will require a mix of local actions and global cooperation. Towards this end, the Departments of Energy in Denmark and U.S. signed a memorandum of understanding (MOU) in 2021 to outline a 5-year framework for cooperation on clean energy research and digitalization science~\cite{DOE_Denmark_MOU_2021}. At the core of this MOU, several areas have been identified for cooperation, including hydrogen technologies, wind energy, energy delivery systems and efficiency, and digital technologies. This paper focuses on describing opportunities and challenges associated with integrating digital technologies and their associated cyber-physical architectures as part of future reliable, sustainable, and affordable energy systems. In particular, we contextualize digitalization in the form of grid flexibility (e.g., large-scale, incentive-based or autonomous control of distributed energy resources or DERs) and discuss architectures that enable scale and how flexibility can engender large-scale, reliable integration of renewables~\cite{Mathieu2025DemandResponse}.

To date, decarbonization has primarily targeted the electric power generation sector, resulting in terawatts of new solar photovoltaics (PV), wind, nuclear, and battery capacity worldwide~\cite{IEA_Electricity_MY2025}. As these variable renewable resources proliferate, aligning electrical demand with the availability of clean supply has become a central challenge for communities and continents alike. At the same time, electrification of heating, cooling, and transportation, including the rapid growth of electric vehicles (EVs), heat pumps, and emerging gigawatt-scale data centers driven by large-language models~\cite{barth2025datacenter}, is reshaping annual load, daily load shapes, and their implications for grid operations, reliability, and cost~\cite{MURPHY2020106878}. In parallel, rapid declines in the cost of internet-of-things (IoT) hardware have made connectivity widely accessible~\cite{Microsoft2019ManufacturingTrends,Chui2021IoT}. As a result, DERs, from appliances and electric vehicles to inverters and behind-the-meter battery systems, can now communicate, share data, and actuate responsively. This ability of DERs to \emph{voluntarily} and \emph{temporarily} adjust their net-power consumption or production in response to external signals engenders \emph{flexibility}.

These developments are fundamentally reshaping the structure and operation of electric power systems. High penetrations of inverter-based resources introduce fast and variable dynamics, while cross-sector electrification creates new inter-dependencies between electricity, heating, transport, and digital infrastructures. Distribution networks are becoming more complex, yet many jurisdictions, particularly in the United States, continue to operate with fragmented data governance, heterogeneous retail market structures, limited observability at the grid edge, and uneven digitalization. Consequently, \textit{digital capabilities have not kept pace with the physical evolution of the grid}, complicating efforts to operate, coordinate, and scale flexible demand-side resources~\cite{Modo2025CurtailmentCrisis}.

Thus, the ongoing energy transition has the potential to produce rapidly growing and large fleets of responsive DERs actuating across different parts of the grid and in pursuit of different objectives. But coordinating these heterogeneous assets reliably and economically is far from straightforward~\cite{almassalkhi2023intelligent}. Without appropriate grid architectures capable of managing interactions between millions of devices, markets, and physical infrastructure, flexibility could lead not to efficiency, but to disorder, creating risks for reliability, resilience, and affordability~\cite{GWAC2024Practical}.

{\color{black}
Recent work has produced powerful methods for characterizing and controlling flexibility at scale, including demand response models and coordination frameworks for large ensembles of DERs~\cite{Mathieu2025DemandResponse}, as well as practical guidance on integration and interoperability challenges for DER coordination in the field~\cite{GWAC2024Practical}. In parallel, grid-architecture efforts have articulated high-level principles for structuring complex electric power systems to accommodate evolving technologies and stakeholders~\cite{taft_2021,matni2024towards}. Yet a persistent gap remains between (i) algorithmic and market-oriented treatments of flexibility and (ii) the cyber-physical \emph{architectures} and interface requirements needed to ensure that flexibility is \emph{reliably admissible} under distribution constraints, dynamic operating conditions, and heterogeneous governance structures. This paper addresses that gap by synthesizing cross-Atlantic lessons (U.S. and Denmark) and by framing flexibility as an architectural design problem that couples infrastructure, data, markets, and control.}

Therein lies the challenge: no clearly defined and validated standards or control architectures yet exist for coordinating DERs in ways that explicitly and systematically account for dynamic grid conditions, distribution constraints, and resilience needs. Moreover, market and regulatory designs differ significantly between Europe (Denmark) and the United States, even though both regions share the same core objectives of reliability, resiliency and efficiency, while also sharing economic and technical principles underlying power system operations and planning. This paper argues that cross-Atlantic collaboration on digitalization and flexibility, which is in the spirit of the Denmark-U.S. MOU, can produce more robust research outcomes for both regions. Scalable and adaptable grid architectures benefit from being developed across diverse systems with heterogeneous actors, objectives, and cyber-physical couplings~\cite{matni2024towards}, much like how universal standards such as USB-C or Bluetooth allow heterogeneous devices to interoperate seamlessly. This makes transatlantic coordination both timely and strategically valuable.

In response to these growing challenges, the U.S. Department of Energy has advanced a series of ``Grid Architecture’’ frameworks that provide a conceptual foundation for designing next-generation power systems~\cite{taft_2021}. These frameworks emphasize two properties as essential for future grid coordination:
\begin{itemize}
    \item \textbf{Adaptivity:} the ability of the grid architecture to adjust to both anticipated and unanticipated changes--including evolving supply mixes, load behaviors, DER proliferation, and cyber-physical disturbances. It requires linking sensing, communication, and control capabilities so that system behavior can be steered in real time as operating conditions change.
    \item \textbf{Scalability:} the capacity of the architecture to integrate an increasingly large number of devices, actors, and interactions without loss of performance or reliability. This entails modular interfaces, interoperable communication pathways, and coordination mechanisms that perform consistently from the device layer to regional or national scales.
\end{itemize}

These concepts motivate the central hypothesis of this paper: achieving reliable and economically efficient coordination of flexible resources is no longer merely a device-level or market-design problem, but a multilevel \emph{architectural} challenge.
{\color{black} Specifically, \emph{we hypothesize that flexibility can be scaled without degrading reliability only when cyber-physical interfaces (data, communications, and control responsibilities) are co-designed with distribution constraints and operating objectives, so that device autonomy can be consistently translated into grid-admissible actions}. We substantiate this hypothesis by (i) contrasting U.S. and Danish infrastructure and governance contexts, (ii) identifying architectural bottlenecks and failure modes that recur across jurisdictions, and (iii) distilling a research agenda with actionable cross-Atlantic collaboration opportunities.
For orientation, Table~\ref{tab:cross_atlantic_snapshot} provides a compact U.S.–Denmark/EU comparison across hierarchy layers; we return to these contrasts throughout Sec.~\ref{sec_currentArchs} and in the challenges and pillars in Sec.~\ref{sec_openChallenges} and~\ref{sec_ResearchAgenda}.}
The remainder of this paper develops this perspective by synthesizing methodological approaches, design principles, and cross-Atlantic lessons that can guide the development of future-ready grid architectures. A graphical overview of the paper's main themes is provided in Fig.~\ref{fig:paperoverview}.

\setcounter{table}{0}
\begin{table}[t]
{\color{black}
\caption{Summary of cross-Atlantic structural similarities \& differences across the hierarchy.}
\label{tab:cross_atlantic_snapshot}
\scriptsize
\setlength{\tabcolsep}{3pt}
\renewcommand{\arraystretch}{1.}
\centering
\begin{tabularx}{\columnwidth}{@{}p{2.125cm}Y@{}}
\toprule
\textbf{Hierarchy layer} & \textbf{Contrast (U.S. vs. DK/EU) $\rightarrow$ architectural implication} \\
\midrule
Transmission and Wholesale &
Nodal ISO/RTO markets see heterogeneous implementations \textit{vs.} zonal markets and national balancing procurement
$\rightarrow$ different price/control signal semantics. \\
\addlinespace
Distribution &
Many state-regulated utilities; uneven flexibility procurement \textit{vs.} clearer DSO mandates and active network management
$\rightarrow$ role clarity shapes interfaces and motivates laminar design. \\
\addlinespace
Data/visibility &
Fragmented data access and privacy with no national backbone \textit{vs.} digitalization and interoperable data spaces
$\rightarrow$ observability and grid-aware dispatch hinges on data pathways. \\
\addlinespace
Aggregator layer &
Strong aggregator ecosystem but utility protocols vary \textit{vs.} more uniform access with structured products demonstrated
$\rightarrow$ need ``minimal interfaces'' between institutions to ensure reliability. \\
\addlinespace
Grid edge &
Split-phase prevalence and DER data asymmetry \textit{vs.} common 3-phase + installer and DSO phase coordination
$\rightarrow$ algorithms may not transfer without interface redesign. \\
\bottomrule
\end{tabularx}
}
\end{table}

In the rest of the paper, we provide an overview of grid fundamentals in Section~\ref{sec_gridFundamentals} to define the physical infrastructure that serves as context for subsequent sections. Section~\ref{sec_flex} introduces flexibility as a system resource and outlines the relevant temporal and spatial scales. Section~\ref{sec_currentArchs} reviews current grid and market architectures in the United States and in Europe, with a particular focus on Denmark. Section~\ref{sec_openChallenges} then describes open technical, systemic, and societal challenges for achieving scalable architectures. Building on these insights, Section~\ref{sec_ResearchAgenda} presents a cross-Atlantic research agenda, including illustrative case studies that identifies priority areas for collaboration to advance a clean and affordable energy transition. The paper concludes in Section~\ref{sec_conclusion} with a summary and call to collaborative action.

\section{Grid Fundamentals and the Physical Infrastructure} \label{sec_gridFundamentals}

\begin{figure}[t]
    \centering
    \includegraphics[width=0.8\linewidth]{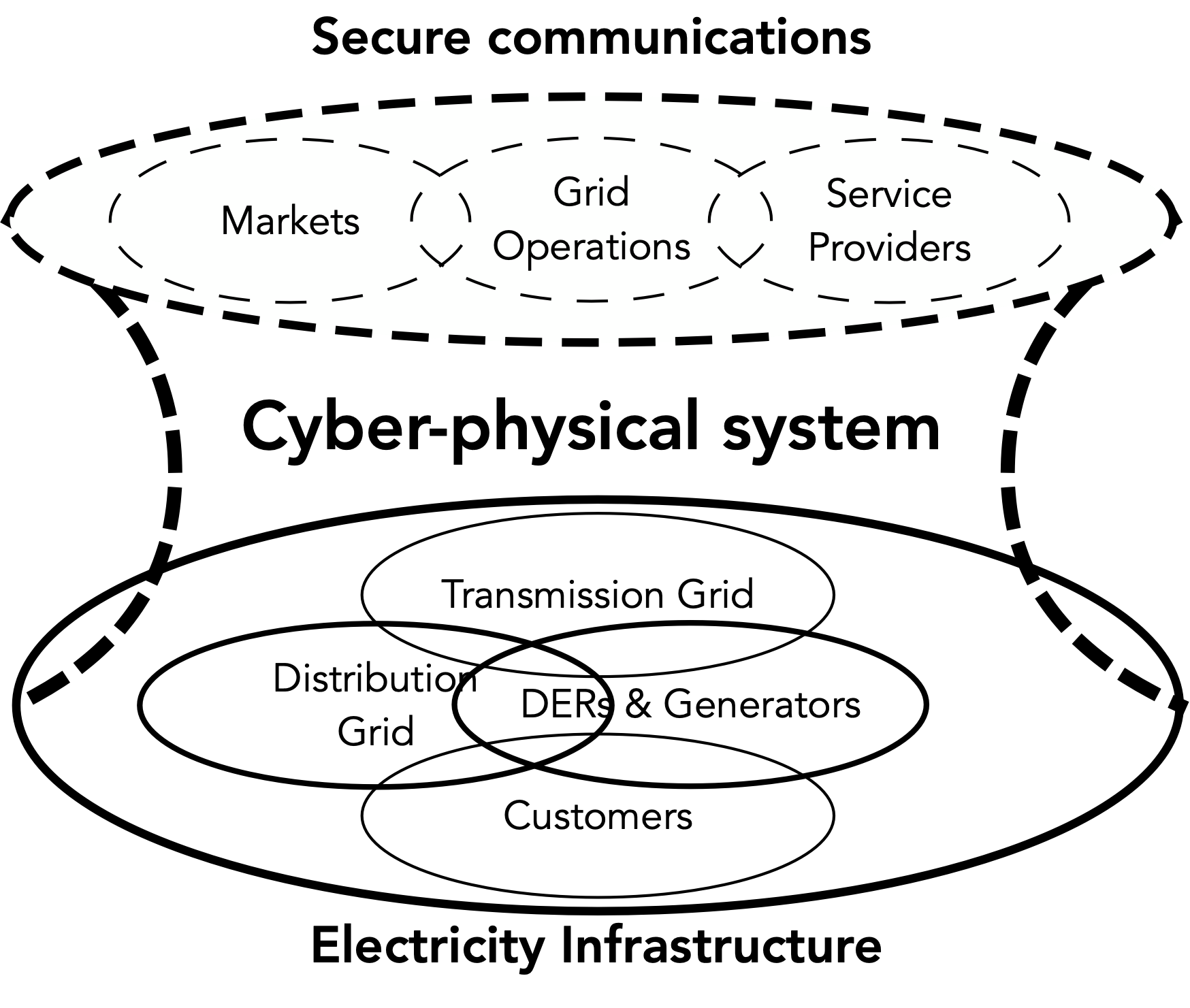}
    \caption{Components of the grid are coupled in a complex cyber-physical system that coordinates data (via secure communications)  and electrons (across electricity infrastructures).}
    \label{fig:hierarchies}
\end{figure}

The electrical grid infrastructure represents a physical (electric) hierarchy made up of numerous energy conversion and storage processes, sensors, and actuators across different voltage layers as illustrated in Fig.~\ref{fig:hierarchies}. The hierarchy effectively has four layers: generation, transmission/sub-transmission, distribution, and customers. Historically, electricity generation was provided almost exclusively by large bulk power plants, which were connected to the transmission or sub-transmission grid at designated buses. This network of interconnected generators created an efficient system for sharing resources and balancing supply and demand over large geographic areas. However, with increasingly distributed generation (DG) and DERs (e.g., solar PV) sited at lower-voltage distribution nodes, the question of how to most efficiently share resources became less clear and gave rise to concepts associated with local energy communities~\cite{NordicEnergy2023,{SwissFederal2023}}, resilience hubs~\cite{Gautam2024GridAware}, and energysheds~\cite{Thomas2021energyshed}. Next, we provide an overview of the electricity infrastructures that interconnect generators, DERs, and customers across the different voltage hierarchies. 
\begin{enumerate}
\item \textbf{Transmission \& sub-transmission grid} (high voltage, generally $>$ 50~kV): transmits energy long distances (at low 1-3\% losses) within a meshed, 3-phase (mostly balanced) network that interconnects regional load centers (e.g., cities) and large multiple-MVA thermal and hydro generators and utility-scale DERs (e.g., wind, battery, or solar PV). At each bus in this network is a transmission substation that steps down voltage to sub-transmission networks (e.g., 345~kV to 69~kV). Sensors, such as remote terminal units (RTUs), have long been integral components of transmission networks. Modern grid-enhancing technologies (GETs) extend beyond sensing to actively increase system capacity and operational flexibility. These include advanced measurement devices (e.g., phasor measurement units, dynamic line rating and conductor temperature sensors), high-performance conductors that raise thermal limits, phase-shifting transformers and optimal power flow controllers that steer real power flows, and topology control mechanisms such as automated switching schemes or modular flexible ac transmission system (FACTS) devices. Collectively, these GETs improve both observability of system states (nodal voltage phasors and current injections) and controllability of power flows, enabling more dynamic, data-driven operation of the transmission grid.
\item \textbf{Distribution grid} (medium voltage, generally $>$ 1~kV): a collection of feeders that are coupled on the secondary side of a distribution substation transformer bank, which steps voltage down (e.g., 69~kV to 13.8~kV) and can host commercial-scale DERs (e.g., 100s of kVA solar PV or batteries). A single distribution feeder is often operated as a radial circuit\footnote{While large urban centers in the U.S. may operate a meshed secondary low-voltage network, this is not the focus herein.} with losses ranging from 2-10\%, where each node typically represents a single-phase pole-top transformer in the U.S. or a 3-phase pad-mount transformer in Europe that steps voltage further down to low voltage (e.g., 13.8~kV to 240~V). 
Distribution feeders may be overhead, underground, or mixed: overhead designs provide easier access and reconfiguration after faults but are more vulnerable to weather and vegetation, whereas underground feeders improve resilience to environmental disturbances at the cost of higher installation expense, slower fault isolation/repair, and more complex impedance profiles that increase modeling and protection complexity. 
Historically, sensing has been limited in distribution systems to just RTUs (on the primary side of substation transformer banks), a few line regulators, and monthly meter readings for billing purposes. The sparse sampling was a reflection of distribution system's historically passive role in energy systems. However, as residential-scale (e.g., $\le$ 25~kVA) DERs are increasingly being deployed in distribution systems, these passive assumptions are rapidly changing in many feeders around the U.S. and Europe.  Since the late 2000s, advanced metering has been deployed at scale across the world and can record individual consumer (net) loads every 1-15 minutes, including active (P) and reactive (Q) power and voltage magnitude. In addition, line sensors and inverters from front-of-the-meter (FTM) DERs can provide time-synchronized data on distribution system states (e.g., voltage phasors). That is, the distribution system is undergoing a digital revolution that can transform distribution utilities from passive load-serving entities to active energy service providers. 
\item \textbf{Customers} (low voltage, generally $<$ 500~V): At the lowest voltage layer, distribution feeders span local suburbs, neighborhoods, or rural towns to supply small commercial businesses and residential home (c\&R) customers.\footnote{Of course, large commercial and industrial (C\&I) customers  may be interconnected at transmission level, such as MVA-scale data centers. However, while important for future grid reliability, their potential flexibility are not the focus of this manuscript.} 
Their electric meter serves as the interface between ``behind-the-meter'' (BTM) kVA-scale DER assets (e.g., roof-top solar PV, EV chargers, other controllable loads) and the FTM distribution grid. The deployment of these BTM assets is partly driven by the proliferation of intelligent electrification programs~\cite{almassalkhi2023intelligent} that is giving rise to aggregated, responsive grid resources often denoted Virtual Power Plants (or VPPs)~\cite{Downing2023VPP}. Note that, in the U.S., it is common for c\&R customers to be interconnected at a single phase (e.g., via a split-phase, center-tapped transformer) while in Denmark, customers are interconnected at all three phases. This difference affects how electrification affects grid reliability (via unbalanced grid operations). In Denmark, distribution network operators (DNOs) coordinate with installers to spread electric heat pumps, EV chargers, or other energy hungry appliances across different phases to support balanced operations, which is an option largely unavailable to U.S. DNOs (utilities), who instead resort to switching consumers to different phases entirely (when that is an option).
Lastly, BTM sensors are increasingly capable of measuring power, voltage, and current at kHz resolution and to wirelessly share these data with service providers and original equipment manufacturers (OEMs) (e.g., Enphase, SolarEdge, and BasePower), enabling more accurate device-level state estimation, real-time demand profiling and forecasting, and enhanced visibility of distribution grids~\cite{CarrieArmel2013Disaggregation,Papageorgiou2025NILMHighFreqReview}.
\end{enumerate}

With grid visibility and controllability increasing across both transmission and distribution systems due to DERs, GETs, and sensors deployments, interesting opportunities arise about how to best coordinate grid-connected assets to ensure a reliable, affordable, and clean energy transition. This cyber-physical coordination will critically depend on $i)$ how market and grid service signals are processed (e.g., via control architectures and algorithms) to modify, shape, or control the individual DERs' behaviors and consequently the aggregate response (moving electrons); and $ii)$ how we can best develop, maintain, and leverage secure and low-cost connectivity and communications (moving data). 

Next, we will highlight the value of flexibility to motivate the need for scalable grid architectures that systematically integrate DERs and enhance grid reliability and decarbonization.
\section{Flexibility as a System Resource}\label{sec_flex}


Flexibility is emerging as a critical enabler of reliable and affordable decarbonization. Yet, flexibility spans heterogeneous devices, diverse objectives, and multiple spatial and temporal scales, which makes it difficult to characterize, value, and operationalize. This section first reviews the system-level benefits of flexibility in Section~\ref{sec:benefits_flex}. It then formalizes the physical attributes that determine what flexibility a resource can provide in Section~\ref{sec:quant_flex}. Finally, Section~\ref{sec:hierarchical_control} describes how flexibility can be realized in practice through distributed and hierarchical coordination schemes that link device behavior to grid and market requirements. Together, these perspectives establish the foundations for the architectural challenges and opportunities that follow in the subsequent sections.

\subsection{Quantifying the benefits of flexibility}\label{sec:benefits_flex}
The benefits of flexibility accrue across multiple temporal and spatial layers of the energy system. At the bulk-power level, flexibility can provide fast reserve capacity, renewable smoothing, and energy arbitrage. At the distribution level, flexibility is often referred to as demand-side flexibility\footnote{Herein, DSF is used interchangeably with flexibility and refers to the ability of end-use customers and DERs to voluntarily and temporarily adjust consumption or production in response to control signals reflecting grid or market conditions~\cite{Mathieu2025DemandResponse}.} (DSF) and can mitigate congestion, avoid or defer capital investments, reduce outage impacts, and support local reliability. When aggregated across millions of devices, flexibility can reduce costs for consumers, utilities, and system operators.

\vspace{2mm}
\noindent\textbf{Multi-market value stack.}  
Across United States wholesale markets, most of the realizable value for distributed batteries and flexible loads comes from capacity-like services rather than real-time energy. Resource adequacy programs, traditional demand-response, and frequency regulation together supply the majority of the revenue stack, while pure energy arbitrage typically contributes a modest share, often below 10 to 25 percent of total value~\cite{Hledik2019LoadFlex,Downing2023VPP,Brattle2025NYGOTF,barth2025datacenter}. This outcome reflects limited real-time price spreads, separation between retail and wholesale prices for behind-the-meter resources, and operational limits on import and export for small aggregated DERs.

A similar pattern appears in several European markets. Although renewable-driven volatility can create occasional arbitrage opportunities, the prevalence of zero-marginal-cost generation often compresses energy margins. As a result, balancing and capacity-like services remain the most stable and scalable revenue streams for distributed flexibility. Examples include frequency containment reserve (FCR), automatic frequency restoration reserve (aFRR), manual frequency restoration reserve (mFRR), and emerging distribution system operator (DSO) flexibility procurement mechanisms~\cite{smarten2022,commission2021,commission2022}. 

\vspace{2mm}
\noindent\textbf{Distribution-level and infrastructure value.}
A major value stream emerges from potential avoided or deferred distribution investments. Recent analyses estimate marginal distribution capacity benefits in New York State of up to \$220/kW-year~\cite{Brattle2025NYGOTF}. European studies similarly project that DSF could avoid between 11–29~billion~EUR of annual low voltage/medium voltage (LV/MV) grid investments through 2030~\cite{smarten2022}, representing 27–80\% of forecasted needs.

\vspace{2mm}
\noindent\textbf{System adequacy and reduced curtailment.}
It is further estimated that by 2030 the EU could face a shortfall of roughly 60~GW of peak capacity; deploying an equivalent amount of DSF would save approximately 2.7~billion~EUR per year relative to new peaking plants~\cite{smarten2022}. DSF can also reduce renewable curtailment substantially, with studies indicating that enabling flexibility could lower annual renewable energy source (RES) curtailment by up to 61\%, thereby improving the economics of wind and solar generation. 
Of course, realizing curtailment savings in practice requires that DSF be electrically proximate to congested renewable generation, since flexibility located far away cannot reliably alleviate local constraints. When combined with a smart coordination architecture, however, this locational dependence becomes an advantage: it naturally couples planning and operations, ensuring that the value of DSF reflects both its control strategy and the strategic siting of flexible resources within the distribution grid.

\begin{figure}[t]
    \centering
    \includegraphics[width=0.7\linewidth]{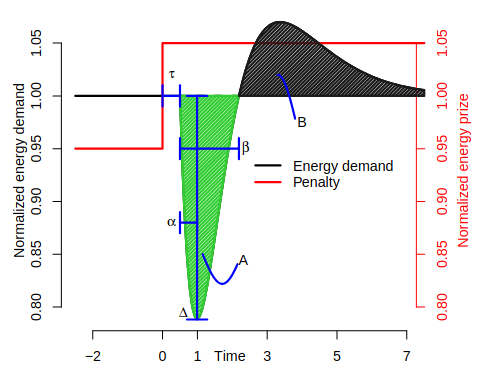}
    \caption{Flexibility Function (FF). The FF links control signals (e.g., prices) to the demand response, providing a standardized description of flexibility. {\color{black}The red line (penalty) shows a step change of the penalty at time $0$. The response starts with a time-delay $\tau$, $\beta$ is the time period with a lower energy demand, and $\alpha$ is the time until the maximum response is reached. The area A represents the reduced load, while the area B is the so-called rebound effect, i.e., the extra load which is needed to compensate for the previously reduced load. The maximum fraction of load reduction is denoted by $\Delta$.}}
    
    \label{fig:Flexibility_Function}
\end{figure}

\vspace{2mm}
\noindent\textbf{Static markets versus dynamic flexibility.}
Although DSF can offer substantial value, existing market designs often reflect \emph{static} clearing logic, for example hourly price volume bids, while flexibility is fundamentally \emph{dynamic}. For instance, a supermarket refrigeration system may respond to a low price in one hour by reducing consumption, but thermal constraints prevent a similar response in the subsequent hour. This basic intertemporal coupling means that many flexible loads cannot express their capabilities via conventional static price-volume bids. Even in organized markets that support multi-interval optimization, such as the California Independent System Operator (CAISO), only resources with explicit energy state models can submit intertemporal bids that reflect state of charge and duration constraints. CAISO implements this through multi-interval real-time dispatch and explicit state of charge feasibility modeling for batteries~\cite{CAISO_DMM_2024}. Similar developments are underway in the Electric Reliability Council of Texas (ERCOT)~\cite{ERCOT_NPRR1204_2023}. 
Recent ERCOT analyses further underscore this challenge: without real-time operational data or functional models for DERs, most flexible loads remain ``invisible'' to system operators and, therefore, cannot participate in dynamic, intertemporal optimization even though their physical behavior is inherently time-coupled~\cite{ERCOT_DER_Operational_Data_2025}.
However, these capabilities are not available for most flexible loads, which cannot submit intertemporal offers that express the coupling between present and future consumption decisions. \textit{As a result, demand-side resources continue to participate as static loads even though their flexibility is inherently dynamic (and stochastic)}. This mismatch highlights the need for interoperability mechanisms, such as Flexibility Functions, that relate market instructions to physical response and allow flexible loads to communicate dynamic capabilities in a more standardized way~\cite{junker2018a}. Figure~\ref{fig:Flexibility_Function} illustrates the intertemporal coupling and how flexibility functions can link control signals (e.g., variables prices or penalties) to demand (e.g., a leaky bucket\footnote{See Buckets, Batteries, and Bakeries taxonomy in Sec.~\ref{sec:quant_flex}.}). See, e.g., \cite{junker2018a} for practical examples.

\vspace{2mm}
In summary, DSF can materially reduce system costs, enhance adequacy, and support decarbonization. But to realize these benefits, grid operators and aggregators require a rigorous framework for describing \emph{what} flexibility a resource can provide and \emph{how} it can respond over time. This motivates the next subsection.

\subsection{Quantifying demand-side flexibility}\label{sec:quant_flex}

Historically, DSF was implemented through coarse mechanisms. Underfrequency load shedding (UFLS) disconnected large blocks of load during emergencies, and traditional demand response (DR) curtailed loads for hours to reduce peaks. These approaches were blunt and infrequent, and they treated demand as largely uncontrollable. Modern DSF is fundamentally different. With pervasive sensing, communication, and networked DERs, demand can now provide short duration, time-coupled flexibility. The magnitude of a deviation is intrinsically linked to its duration, reflecting thermal constraints, comfort preferences, or process requirements. These dynamics differ substantially from those of traditional generators that historically supplied balancing services~\cite{kundur1994power}. As a result, realizing DSF in practice requires distributed coordination schemes that explicitly model intertemporal constraints and respect the cyber physical behavior of loads. This motivates the need for hierarchical and device-aware control architectures that can transform heterogeneous DERs into reliable, aggregate system resources.

\vspace{2mm}
\noindent\textbf{Core physical characteristics.}
To design such coordination schemes, it is necessary to describe DSF in terms of the underlying physical limits of the devices that supply it. Regardless of whether the resource is a heat pump, an EV charger, a refrigeration system, or a commercial process, its flexible behavior can be characterized using four key attributes:
\begin{itemize}
    \item \emph{Power limits} (MW): bounds on instantaneous upward or downward deviation.
    \item \emph{Energy limits} (MWh): cumulative bounds reflecting thermal inertia, state of charge, or process requirements.
    \item \emph{Ramp rate limits} (MW/s): effective limits induced by communication, sensing, or coordination latency, even when devices themselves can switch rapidly.
    \item \emph{State of energy} (MWh): time-varying internal state (temperature, state-of-charge, process stage) that constrains future capability.
\end{itemize}

These characteristics are time varying and stochastic due to occupant behavior, weather uncertainty, device heterogeneity, and modeling error. DSF from aggregated loads is often modeled using stochastic differential equations~\cite{malhame_loadmodel_1985,madsen2007a} or mean field approximations~\cite{mathieu_state_2013,chen_individual_2014,PEMqos2020tpwrs}.

\vspace{2mm}
\noindent\textbf{Taxonomy: Buckets, Batteries, and Bakeries.}
A useful classification introduced in~\cite{PEHBS:tfmceadsg} partitions flexible loads into:
\begin{itemize}
    \item \textbf{Buckets:} energy-limited resources that ``leak'' (e.g., building heating/cooling with thermal loss).
    \item \textbf{Batteries:} resources requiring a targeted future energy level (e.g., EV charging before departure).
    \item \textbf{Bakeries:} process-driven loads that must run for a minimum duration once activated (e.g., dishwashers, dryers).
\end{itemize}
This taxonomy highlights the diversity of constraints and helps identify feasible flexibility shapes.

\vspace{2mm}
\noindent\textbf{Ramp-rate as an emergent system property.}
Although individual DERs may switch or modulate power almost instantaneously, the \emph{aggregate} ramp-rate of a coordinated fleet is fundamentally limited by the cyber–physical coordination loop.  
For example, the sensing–communication–computation-response cycle takes 1~second, a fleet of \(N\) identical loads of capacity \(\bar P\) kW can ramp at most \(N\bar P\)~kW/s, if all devices are actuated simultaneously (e.g., via a broadcast signal), or only \(\bar P\)~kW/s if devices must be switched sequentially. Thus, ramp-rate is not merely a device attribute; it is a \emph{property of the control architecture}.  
This point motivates the architectural considerations developed in Section~\ref{sec_openChallenges}.

\vspace{2mm}
\noindent\textbf{Estimating and forecasting states.}
To coordinate DSF at scale, aggregators must estimate the internal ``energy state'' of flexible loads using physics-based models, Kalman filtering~\cite{meyn_ancillary_2015,PEMsysProp2018cdc}, or data-driven methods~\cite{matar2023learning}. Accurate state estimation underpins reliable forecasting and optimal dispatch of available flexibility and expected performance.

\vspace{2mm}
Together, these attributes form the basis for modeling, quantifying, and ultimately coordinating DSF. They also highlight why DSF requires architectural approaches that unify device-level dynamics, grid constraints, and market interactions, which are topics developed in the next section.

\subsection{Realizing demand-side flexibility with distributed control}\label{sec:hierarchical_control}

The physical characteristics in Section~\ref{sec:quant_flex} must be translated into system-level actions that support balancing, congestion management, and resource adequacy. This requires coordination schemes that link local device objectives with grid objectives while respecting privacy, autonomy, and device constraints.

\vspace{2mm}
\noindent\textbf{From heterogeneous devices to coordinated system response.}
Individual flexible loads follow their own thermal, electrochemical, or process dynamics, which determine how long a deviation from baseline can be sustained. To provide reliable system services, aggregators must therefore coordinate many heterogeneous devices so that the aggregate response is predictable, stable, and dispatchable. This coordination challenge couples device physics, communication latencies, and market signals, and it cannot be addressed using traditional market constructs or device-agnostic control signals alone. Intertemporal constraints, stochastic availability, and cyber–physical delays must be explicitly represented. These requirements motivate standardized interfaces and hierarchical control architectures that translate system-level objectives into device-feasible actions while preserving privacy and autonomy. The remainder of this subsection introduces one such approach based on Flexibility Functions.

\vspace{2mm}
\noindent\textbf{Hierarchical control and Flexibility Functions.}

\begin{figure}[t]
\centering
\includegraphics[width=0.7\linewidth]{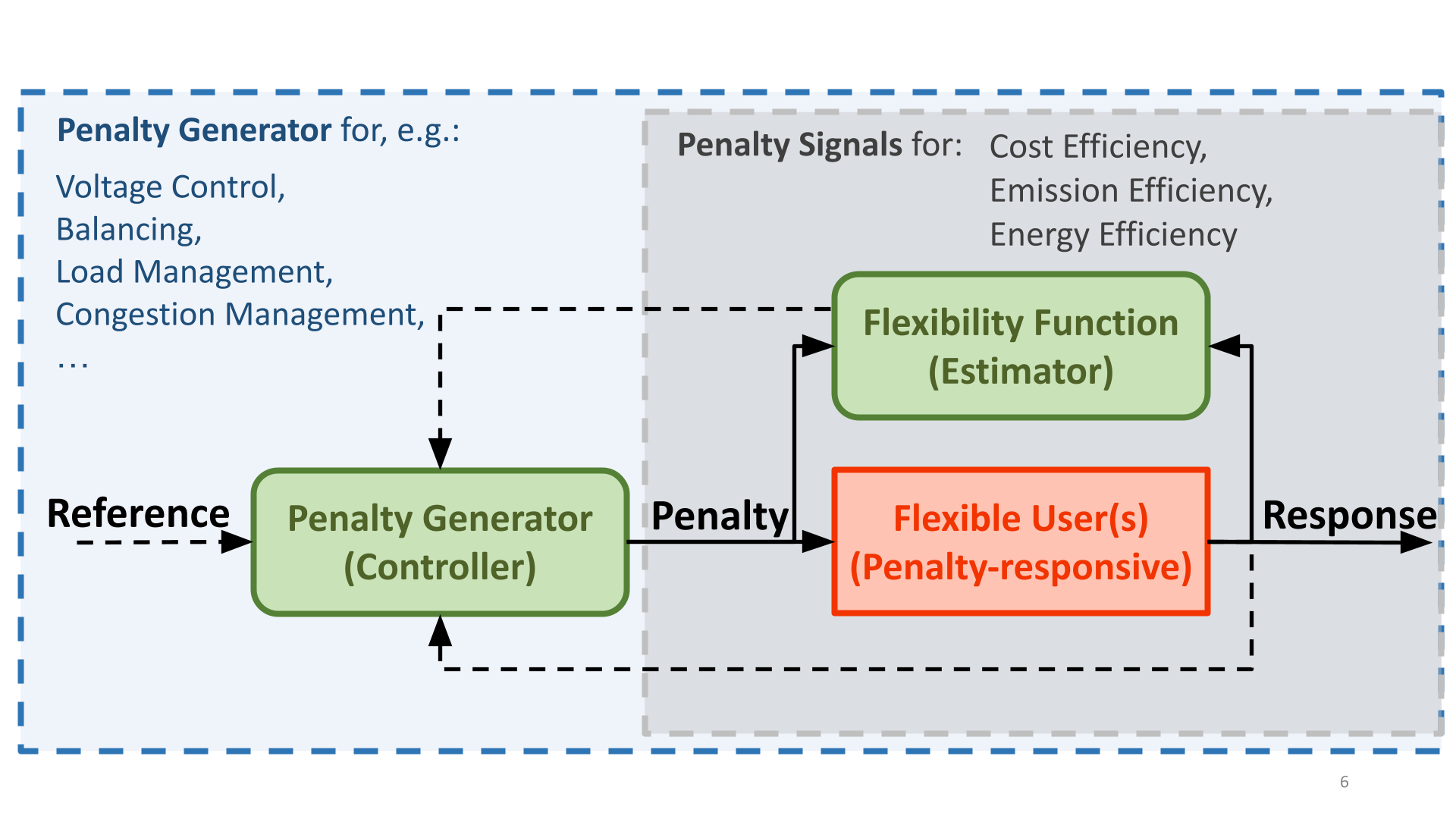}
\caption{Hierarchical control for utilizing demand-side flexibility (DSF).}
\label{fig:flexible_energy}
\end{figure}

A practical pathway for realizing DSF at scale is to employ hierarchical control architectures in which upper layers broadcast incentive or activation signals and lower layers interpret these signals through device-level controllers. As shown in Figure~\ref{fig:flexible_energy}, the upper-level controller issues a time-varying control input (e.g., a dynamic price or grid-service request), while each device autonomously adjusts its demand according to its internal state, constraints, and local objectives; the resulting ensemble response yields an aggregated flexibility profile for balancing, congestion management, and other services. {\color{black} To expose this behavior through a standardized, technology-agnostic interface, we can use \emph{Flexibility Functions (FFs)}~\cite{junker2018a}, which map an external signal trajectory to an expected power/energy response over time, including dynamics, time-coupling, and rebound effects. For example, a thermostatically controlled load’s FF can predict that a 15-minute price spike produces a temporary load reduction during the event and a short rebound afterward as the device restores its temperature, whose effect is also illustrated in Fig.~\ref{fig:Flexibility_Function}.}

\begin{figure}[t]
    \centering
    \includegraphics[
    width=0.95\linewidth,
    trim={0in 0in 2.5in 0in},
    clip
    ]{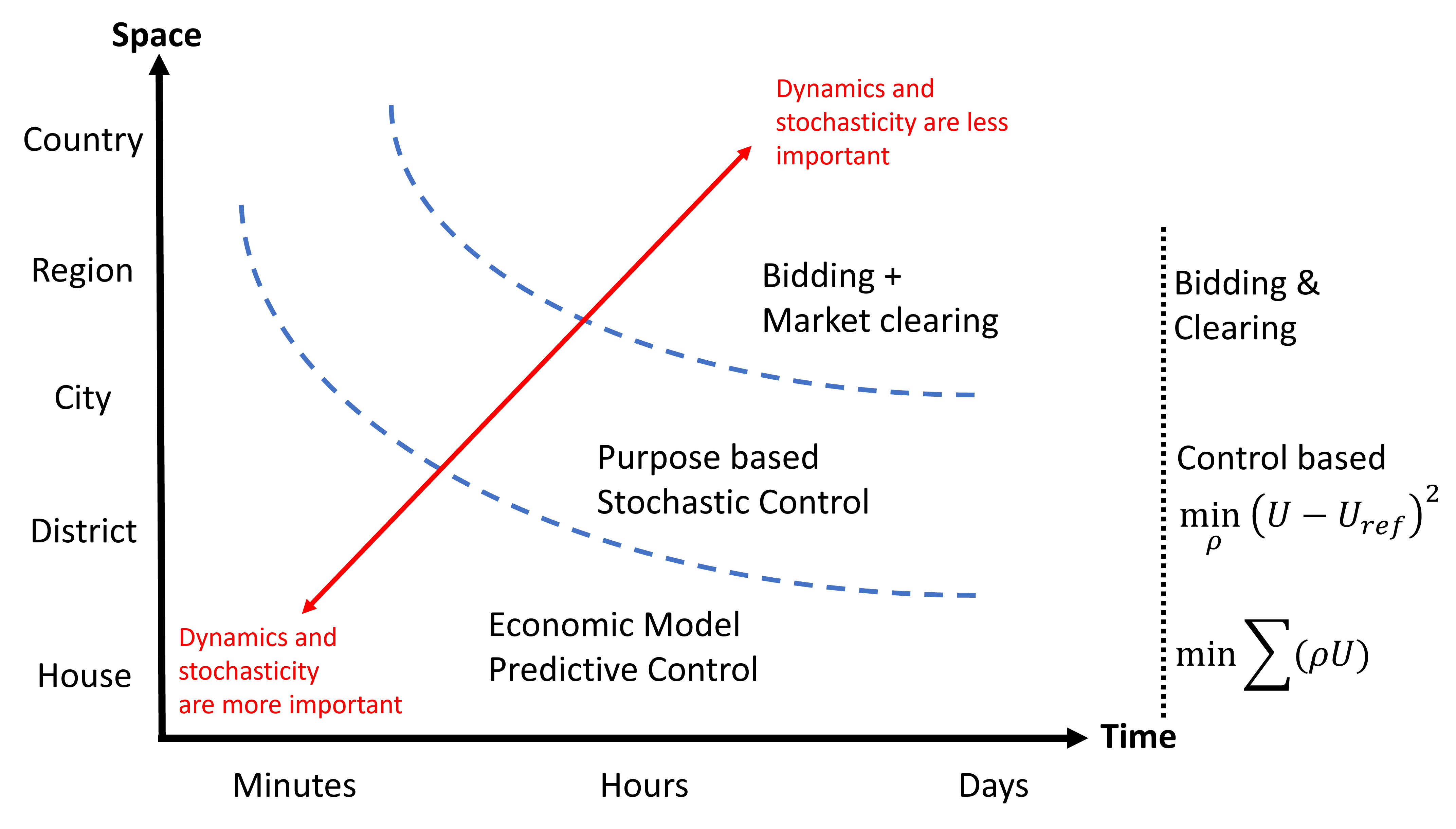}
    \caption{
    Overview of control and market interactions across different time and spatial scales and the effects of dynamics and stochasticity. {\color{black}At the slower timescales and for larger aggregations of resources, conventional market ``bidding and clearing'' is most relevant and dynamics and stochasticity are less important. At finer spatial or temporal scales, control-based approaches explicitly leverage dynamics and stochasticity when activating flexibility to realize stochastic or predictive controllers. 
    }
    }
    \label{fig:control_hierarchy}
\end{figure}

FFs can be identified using data-driven or physics-based models and then incorporated into hierarchical control schemes. At the device level, controllers optimize local objectives such as comfort, cost, or process requirements. At the upper level, an aggregator uses FFs to schedule and allocate flexibility across large portfolios to support peak shaving, balancing, voltage management, and other grid services. Because FFs abstract device-specific behavior, they preserve autonomy and privacy while enabling coordination at different scales, as illustrated in Fig.~\ref{fig:control_hierarchy}.

These properties have motivated ongoing work in Europe to formalize FFs as a Minimum Interoperability Mechanism~\cite{EU_OpenDEI_2022}. Although FFs do not replace detailed device models, they provide a common language (i.e., interface) for representing flexibility in a scalable and technology-agnostic manner. This aligns with the broader architectural themes developed in Section~\ref{sec_currentArchs}.

\section{Current Grid and Market Architectures} \label{sec_currentArchs}

This section summarizes how grid and market architectures in the United States and Europe evolved from generator-centric coordination to the present landscape, where DERs increasingly shape operational and planning decisions. The two regions have largely shared common decarbonization goals, but differ significantly in institutional design, market structures, data governance, and the roles (and responsibilities) assigned to system operators. These differences shape the opportunities and constraints for deploying DSF at scale.

\subsection{United States: Nodal markets and heterogeneous retail structures}

Today, the United States operates a highly decentralized institutional landscape with multi-layer control loops across the transmission operator (ISO/RTO), distribution utilities (DSOs), aggregators, and BTM DERs. {\color{black} Wholesale markets generate system-level control and price signals. Distribution utilities enforce grid constraints and operational limits regionally. DER coordinators translate system signals into device instructions and the physical devices self-dispatch according to local objectives (e.g., need for energy, tariffs). Feedback flows upward through telemetry, metering, and state estimation, completing the cyber-physical control loop that governs real-time grid and market interactions~\cite{taft2021gridarch}.}

Bulk-power coordination is carried out by regional transmission organizations and independent system operators (RTOs/ISOs), which operate wholesale markets and coordinate system reliability under federal oversight by the Federal Energy Regulatory Commission (FERC), with reliability standards developed by the North American Electric Reliability Corporation (NERC). {\color{black} Examples include PJM, ISO New England (which serves Vermont), the New York ISO (NYISO), the Midcontinent ISO (MISO), the California ISO (CAISO), and ERCOT in Texas. In Vermont, the high-voltage transmission system is owned and operated by Vermont Electric Power Company (VELCO), a statewide transmission-only utility. These market rules and reliability processes differ across regions and materially affect how flexibility can be offered, scheduled, measured, and compensated.}


At the distribution level, roughly three thousand utilities operate under diverse retail structures and regulatory jurisdictions, each set by individual states. Consequently, retail tariffs, interconnection standards, telemetry requirements, and aggregator participation rules vary widely, creating a fragmented policy landscape. DSF providers may thrive in one state yet struggle in another because DER markets are shaped locally: California rewards behind-the-meter flexibility through the Demand Response Auction Mechanism (DRAM), resource-adequacy procurement, and high rooftop-solar penetration, enabling models like OhmConnect; Texas’s ERCOT market favors merchant economics, driving rapid growth of utility-scale batteries and PV but offering little standardized support for residential aggregation; New York sits between these extremes with ambitious targets and layered oversight, but slow early deployment and New York City battery restrictions have raised transaction costs. With tariff structures (time-of-use, demand charges, export credits) and interconnection rules defined locally, what works in California often does not translate to Texas or New York~\cite{Modo2025CurtailmentCrisis,Miller2025_CAvsTX_BESS}.

Data governance is similarly heterogeneous. State-level advanced metering rollouts and utility data access programs differ widely, with Vermont at $>95\%$ while New York State is around $70\%$. This also limits visibility of loads, DERs, and local network conditions for many system operators. Although initiatives such as the Green Button standard and state-level data platforms, such as, the New York State Energy Research and Development Authority DER database\footnote{NYSERDA maintains an open database of DERs at \url{https://der.nyserda.ny.gov/}.} (NYSERDA), exist in isolated cases, there is no national data backbone for distribution-level coordination.


Participation of flexible demand in organized U.S. wholesale markets is evolving. FERC Order~2222 requires RTOs and ISOs to enable aggregated DERs to participate directly in wholesale markets, but implementation timelines, telemetry requirements, and coordination protocols vary significantly across regions. Offer structures remain largely static and do not support expression of intertemporal constraints for flexible loads, as discussed in Section~\ref{sec:quant_flex}. As a result, most demand-side flexibility still participates indirectly through aggregators or through utility-run peak management programs.

\begin{figure}[t]
    \centering
    \includegraphics[width=0.8\linewidth]{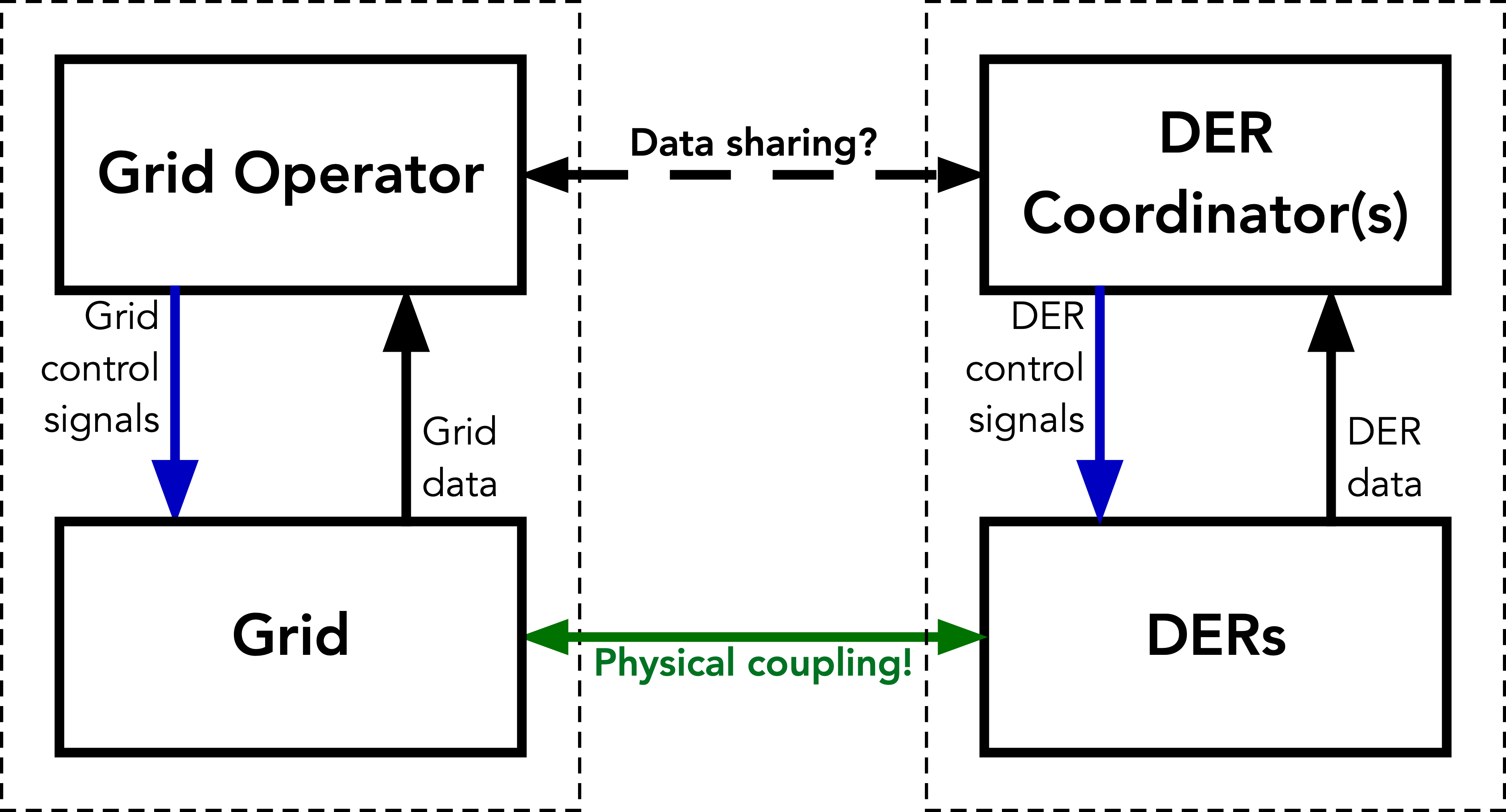}
    \caption{Grid-aware coordination requires clearly defined roles and data pathways. DER operators typically have detailed device-level information but limited visibility into distribution grid conditions, while utilities observe portions of the grid but have limited access to behind-the-meter assets.}
    \label{fig:gridaware}
\end{figure}

At the same time, increasing DER activity introduces operational complexities for utilities. Under Order~2222, distribution utilities may override or block DER dispatches that they determine would compromise local reliability or cyber-security, but the Order does not specify how utilities should make such determinations or what data and analytic capabilities are required. In practice, these decisions are often tied to coarse hosting-capacity maps (updated annually) or emerging ``flexible interconnection'' processes, where utilities provide limited local grid data to inform allowable DER operating envelopes. These mechanisms highlight the broader data-sharing and coordination challenges between utilities, DER operators, and market participants. Figure~\ref{fig:gridaware} illustrates this information asymmetry: DER operators typically have device-level visibility but lack grid context, while utilities have partial grid visibility but limited access to behind-the-meter asset states. This asymmetry is a central barrier to scaling DSF under the current U.S. architectural model.

\subsection{Europe and Denmark: Zonal markets, structured roles, and digitalization}

European electricity systems operate under a different institutional paradigm. Energy markets are organized zonally, with coordinated day-ahead and intraday markets operated by Nord Pool and other regional exchanges. Ancillary services are procured nationally and increasingly include explicit products for fast frequency response and mFRR. Market coupling across borders has progressed significantly, creating joint clearing and coordinated transmission scheduling.

Distribution system operators in Europe have clearer statutory roles than in the United States. They are mandated to ensure local reliability, enable DER integration, publish long-term grid plans (similar to Integrated Resource Plans in the U.S.), and develop active network management capabilities. Many DSOs in Denmark, the Netherlands, Germany, and the UK conduct pilots in congestion management, flexibility procurement (often via local markets), and local balancing. This structured role shapes how DSF can be activated at local levels.

Data governance and digitalization efforts are significantly more coordinated. Denmark’s Energinet operates a national DataHub for electricity metering data. In parallel, Center Denmark leads multi-energy data platforms that integrate electricity and heat system data across hundreds of utilities. At the EU level, data spaces, interoperability frameworks, and cross-sector integration initiatives aim to create standardized digital infrastructures that support sector coupling and flexibility activation.

Market access for flexible demand is also more uniform. European balancing markets accept aggregated load and distributed resources subject to telemetry and verification requirements. Although local flexibility markets remain in pilot stages, DSOs across Denmark, the UK, and the Netherlands have implemented demonstration projects for congestion management and local balancing.

\subsection{Summary}

The United States offers a rich experimental landscape with innovative technology companies (aggregators), diverse utility practices, and sophisticated nodal markets. Europe and Denmark provide structured DSO roles, coordinated data infrastructures, and growing integration of electricity and heating sectors. These differences create complementary opportunities for DSF research. Understanding how flexibility can be activated under each paradigm motivates the architectural challenges described next.

\section{Open Challenges for Scalable Architectures}
\label{sec_openChallenges}

Millions of DER devices are now installed across diverse utility territories and aggregator portfolios, each operating under different retail tariffs, interconnection rules, and market designs. This creates a cyber–physical–economic system, whose behavior is shaped not only by device physics but also by communication delays, market (or control) design, institutional roles, and consumer adoption. 
{\color{black} The U.S. Department of Energy’s grid architecture work emphasizes that future distribution grids must embody \emph{adaptivity} and \emph{scalability}~\cite{taft_2021}.  Accordingly, the remainder of this section organizes the resulting open challenges into \emph{technical}, \emph{systemic}, and \emph{societal} categories, building on the insights in Sections~\ref{sec:quant_flex} and~\ref{sec_currentArchs}.

\noindent\textit{Scope note:} In this section, ``Europe/EU'' denotes a family of institutional patterns rather than a uniform baseline; implementation of smart metering, retail price flexibility procurement varies substantially across EU member states, just as U.S. practice varies across ISOs/RTOs, states, and utilities.}


\subsection{Technical challenges}
Realizing demand-side flexibility at scale requires reliable information flows, validated models, and secure cyber–physical integration. Several technical gaps remain:
\begin{itemize}
      \item \textbf{Observability and telemetry:}
    Distribution systems were historically operated with very limited visibility: few synchronized sensors, sparse supervisory control and data acquisition (SCADA) coverage, infrequent advanced metering infrastructure (AMI) readings, and incomplete knowledge of feeder topology or device status. Operating ``in the dark'' necessitated distribution operators to rely on large engineering safety margins and conservative planning practices, which in turn led to today's underutilized grid and a bias toward capital investment rather than operational flexibility. Modern DSF reverses this logic: using flexibility to solve local constraints requires accurate, time-stamped knowledge of voltages, flows, and device states. Scalable architectures therefore depend on distribution system state estimation (DSSE), updated meter data, topology and parameter identification, and integration of AMI~2.0 and hybrid protection-and-sensor streams~\cite{Geth2023DSEPractical}. Without sufficient observability, grid-aware dispatch of DERs cannot enforce constraints reliably, while settlement, baseline construction, and long-horizon planning remain fragile. The challenge is architectural as much as technical: how to design interoperable pathways for streaming grid data, DER telemetry, and state estimates so that planning and operations can be jointly informed rather than siloed.
    
    \item \textbf{Interoperability:}
    Modern DER fleets expose capabilities through heterogeneous and partially overlapping communication standards, including OpenADR, IEEE~1547, IEC~61850, SunSpec Modbus, and emerging interoperability profiles from the National Institute of Standards and Technology (NIST). NIST has developed a framework and roadmap that defines ``communication pathway scenarios''~\cite{gopstein_2021}, outlining where information exchange must occur between devices, aggregators, utilities, and markets. In high-DER scenarios, dynamic attributes such as ramp-rates, intertemporal limits, and grid constraints must be communicated reliably across these pathways. Interoperability profiles, such as mappings between IEEE~1547 interconnection requirements and IEC~61850 device models~\cite{oral_2022}, provide a structured way to reduce implementation complexity, but remain incomplete and unevenly adopted. As inverter functionality evolves (e.g., toward grid-forming behavior), these profiles must adapt to ensure that targeted system attributes remain achievable at scale.

    \item \textbf{Data access and sharing:}
    Utilities and aggregators operate with asymmetric visibility: aggregators have detailed device-level telemetry but little grid context, while utilities observe only portions of the network and rarely receive behind-the-meter states. Proprietary application programming interfaces (APIs), inconsistent time-stamping, differing data models, and privacy constraints further hinder coordination. NIST’s communication-pathway scenarios identify these data gaps explicitly, emphasizing the need for interoperable data schemas, role-based access, and secure low-latency pathways that support both real-time operations (e.g., FlexRequest/FlexOrder signaling) and planning tasks such as DSSE, topology identification, and baseline verification. Without standardized, bidirectional data access, architectures cannot effectively couple DSF dispatch with grid reliability constraints.

    \item \textbf{Cyber-resilience:}
    As DSF scales, millions of internet-connected thermostats, EV chargers, batteries, and inverters expand the grid’s attack surface. Device-level compromises can propagate through aggregators, cloud services, or distribution automation, creating systemic risks. Scalable architectures therefore require secure-by-design device interfaces, authenticated command pathways, encrypted firmware updates, and anomaly detection that spans both cyber and power-system signals. NIST’s roadmap identifies these as critical requirements for high-DER communication pathways: interoperability must be achieved without sacrificing security, and cyber-physical co-simulation is needed to validate that DSF coordination does not introduce new vulnerabilities.

    \item \textbf{Measurement and verification (M\&V):}
    Baseline estimation and counterfactual construction remain challenging for heterogeneous, time-coupled loads. For thermostatic and process-driven resources, rebound effects and intertemporal constraints make baseline drift common, and standard regression-based methods often miss these dynamics. Even for well-instrumented resources such as EV chargers and batteries with high-resolution telemetry, irregular user behavior, shifting state-of-charge (SOC) targets, and opportunistic charging complicate the construction of reliable counterfactuals for settlement. As a result, M\&V for DSF remains fragile in many practical settings and is a limiting factor for scaling flexible load participation.

    \item \textbf{Modeling and validation:}
    High-fidelity digital twins, hardware-in-the-loop testbeds, and co-simulation frameworks that integrate communication delays, device heterogeneity, and distribution dynamics are still emerging.

    \item \textbf{Stability and synchronization:}
    Because DSF is inherently state-dependent and intertemporally coupled, as characterized in Section~\ref{sec:quant_flex}, broadcast (price or control) signals can drive large fleets of DERs into synchronization. This synchronization can amplify rebound effects, reduce controllability, or even destabilize the grid. Scalable DSF architectures, therefore, require local safeguards, diversity in incentives, and explicit consideration of aggregated dynamic responses.
\end{itemize}

\begin{remark}[Cross-Atlantic Context]
    {\color{black} Denmark and several leading European pilots often operate in well-instrumented environments (e.g., unified telemetry requirements, interoperability profiles, and three-phase residential supply), which can simplify validation of DSF coordination schemes at scale; however, these capabilities remain heterogeneous across EU member states.}
    In the United States, household service is typically single split-phase and telemetry or interconnection rules vary by state and utility, leading to greater heterogeneity in device behavior, data quality, and operational oversight. Although these structural differences pose challenges, they also create opportunities for co-learning: European pilots offer well-instrumented environments for testing standardized architectures, while the diverse and fragmented U.S. landscape provides real-world stress tests for scalability, cyber–physical robustness, and market integration. Joint cross-Atlantic efforts can, therefore, use these complementary contexts to design architectures that are both interoperable and resilient across fundamentally different grid structures.
\end{remark}

\subsection{Systemic challenges}
Even with robust technical tools, grid architecture must reconcile sectoral coupling, institutional roles, and heterogeneous market structures.
\begin{itemize}
\item \textbf{Sector coupling:}
    Electrification of heating, cooling, transport, and industry links previously independent infrastructures through electricity. This improves efficiency but reduces diversity of energy carriers, creating new reliability risks. When heating, mobility, and industrial processes all depend on the electric grid, disturbances that were once localized can cascade across sectors, increasing the potential for highly correlated failures. In parallel, emerging all-electric loads such as data centers, electrolyzers for hydrogen, and power-to-X facilities introduce large, dynamic, and often weather-correlated demand whose interactions with distribution networks remain difficult to model~\cite{smillie2023gaselectric,poulsen2025LCA_P2X,twitchell2023ldes,Sepulveda2021LDES,chen2025electricity}. These interdependencies complicate reliability analysis, contingency planning, and adequacy studies, since stress events now propagate across coupled sectors rather than remaining isolated within one infrastructure.

    \item \textbf{Transmission-distribution operator coordination:}
    Roles, incentives, and data flows between transmission operators and distribution utilities remain unsettled. Without clear coordination protocols, co-optimization across voltage levels remains challenging.

    \item \textbf{Market heterogeneity:}
    Nodal markets in the United States and zonal markets in Europe lead to different procurement mechanisms, pricing signals, and flexibility needs. This complicates scaling solutions across regions and limits common tooling. {\color{black} In Europe, day-ahead and intraday coupling is implemented through common network codes and shared clearing mechanisms across bidding zones, whereas U.S. ISO/RTO markets differ in product definitions, settlement, and coordination rules—differences that shape how flexibility can be valued and activated across jurisdictions.}

    \item \textbf{Distribution-level procurement:}
    Unlike wholesale markets, distribution systems lack mature incentive mechanisms for accessing flexibility. Local flexibility markets have shown promise in European pilots~\cite{smartnet,tsaousoglou2021mechanism}, but often suffer from thin bidder pools and complex bidding formats~\cite{madina2019a}.

    \item \textbf{Mismatch between static markets and dynamic flexibility:}
    As discussed in Section~\ref{sec:quant_flex}, many flexible assets have intertemporal constraints: \emph{a deviation now affects feasible deviations later}. Static price–volume bids cannot express these dynamics. Aggregator-mediated approaches, where an entity trades in wholesale markets and broadcasts dynamic prices or activation signals, have shown promise (e.g., up to 50\% savings in wastewater treatment plants~\cite{stentoft2021a}), but lack standardized interfaces and regulatory incentive mechanisms.

\end{itemize}

\begin{remark}[Cross-Atlantic Context]
U.S. nodal markets and diverse retail arrangements make distribution coordination challenging. 
{\color{black} Denmark and several EU member states offer clearer DSO mandates and more coherent data governance, but distribution-level procurement mechanisms and retail arrangements remain heterogeneous across Europe. At the wholesale layer, EU market coupling and network codes drive greater cross-border harmonization, while U.S. market products and coordination protocols are implemented through ISO/RTO-specific tariffs.}
These differences also present architectural opportunities. Danish pilots can help U.S. utilities test coordinated DSO–aggregator interfaces, congestion management schemes, and Flexibility Function-based coordination under controlled settings before translation to heterogeneous U.S. environments. Conversely, the scale, diversity, and device heterogeneity of U.S. pilots provide stress tests that can inform European architectural robustness and help refine flexibility procurement rules for larger and more variable populations. This cross-learning potential makes trans-Atlantic pilot adaptation a practical pathway for accelerating progress on scalable architectures.
\end{remark}

\subsection{Societal challenges}

Ultimately, demand-side grid flexibility depends on people: consumers, installers, aggregators, utilities, and regulators. Several societal challenges currently limit scale:

\begin{itemize}

\item \textbf{Consumer adoption and workforce:}
Flexibility programs must deliver visible value with minimal effort. Traditional utility-led programs have often faced slow uptake due to enrollment friction, limited automation, and unclear incentives. Technology-driven DER coordinators and aggregators, by contrast, increasingly excel at on-boarding customers through software, apps, and turnkey device integration. Scalable adoption requires architectures in which utilities and aggregators cooperate rather than compete. Utilities must provide grid-aware operating envelopes and reliability oversight, while aggregators deliver consumer-facing automation, engagement, and service quality. Workforce constraints further limit deployment. Shortages of electricians, heating, ventilation, and air conditioning (HVAC) technicians, and DER installers slow adoption even in well-funded programs.

\item \textbf{Equity:}
DSF benefits often accrue disproportionately to affluent early adopters who can afford DERs and who are frequently grandfathered into favorable rates or incentive structures. As adoption rises, and in the absence of scalable architectures for coordinating DERs, system costs can increase due to back-feed constraints, new infrastructure needs, and reduced volumetric revenues. These costs are often shifted to late adopters and non-adopters through higher distribution charges, magnifying affordability challenges and widening energy inequities. Equity-centered cost-allocation and program-design frameworks remain limited.

\item \textbf{Privacy and data governance:}
State-level rules in the United States differ widely. Many consumers resist sharing granular usage data, and durable frameworks for consent, data rights, and federated data-sharing remain nascent. This uncertainty limits both DSF participation and regulator ability to evaluate program performance.

\item \textbf{Integration costs:}
Many device APIs incur annual per-device fees from OEMs and offer limited telemetry or actuation capabilities. These limits constrain services faster than 15~minutes and erode the economics of DSF portfolios.

\end{itemize}

\begin{remark}[Cross-Atlantic Context]
{\color{black} During the 2022 energy crisis, several European markets demonstrated rapid consumer responsiveness to price signals. Danish households, in particular, adjusted heating and appliance use within days of sharp price increases~\cite{Madsen2024EnergyVulnerability,Bak2025EnergyCrisis}.} This led to broad public acceptance and DSO implementation of time-of-use rates in Denmark, which caused a measurable reshaping of load profiles and, most notably, a new clustering of demand just after midnight when the low-price period begins~\cite{Randewijk2024}. This experience demonstrates how clear price signals, standardized digital infrastructure, and strong consumer trust can activate flexibility rapidly. {\color{black} However, access to smart meters and dynamic retail pricing remains uneven across EU member states.}

The United States faces different constraints. State-by-state privacy rules, heterogeneous retail tariffs, single-phase residential supply, and limited distribution level telemetry complicate both enrollment and reliable activation of DSF. However, these differences create opportunities for mutual learning. Danish and EU pilots benefit from three-phase supply, national data hubs, and consistent telemetry standards. U.S. pilots operate under more fragmented rules, but they provide testbeds for architectures that must function without centralized data platforms.

For regulators and system operators on both sides of the Atlantic, the shared lesson is that scalable DSF requires trustworthy telemetry, standardized interfaces, transparent data flows, and clear consumer protections. Cross-Atlantic collaboration can help regulators make data-informed decisions that safeguard affordability, reduce inequities, and enable flexibility programs that scale.
\end{remark}

\subsection{Summary}

The technical, systemic, and societal challenges outlined above illustrate why scalable architecture for demand-side flexibility cannot be achieved through device-level innovation alone. Progress requires coordinated advances in grid visibility and telemetry, interoperable communication pathways, cyber–physical control, and measurement and verification, but it also depends on institutional design, market mechanisms, consumer engagement, and equitable cost allocation. Taken together, these challenges define an architectural gap: today’s systems lack the mechanisms needed to translate static market constructs and fragmented grid data into dynamic, device-feasible actions at scale. Bridging this gap requires a unifying framework that connects device physics, coordination architectures, and market signals across the U.S. and European contexts. Section~\ref{sec_ResearchAgenda} outlines a cross-Atlantic research agenda for developing such a framework, emphasizing Flexibility Functions, Minimum Interoperability Mechanisms, coordinated testbeds, and regulatory co-design as key pathways toward scalable and adaptive distribution-level flexibility.

\section{Cross-Atlantic Research Agenda on Scalable Architectures} \label{sec_ResearchAgenda}



The challenges identified in Section~\ref{sec_openChallenges} make clear that scalable demand-side flexibility cannot be achieved through isolated advances in technology, markets, or policy. These challenges reveal an architectural gap: today’s systems lack the cyber-physical, data, and market interfaces needed to translate device-level flexibility into reliable grid-level actions at scale.
 Instead, progress requires coordinated innovation across cyber-physical control, data and interoperability standards, digital verification tools, and regulatory and market design. The United States, Denmark, and the broader European Union are uniquely positioned to pursue such a joint agenda. Their complementary strengths include the scale and heterogeneity of U.S. DER ownership models and aggregator ecosystems, the structured DSO roles and advanced multi-energy digitalization initiatives in Denmark and Europe, and the availability of accessible national testbeds and data infrastructures across both regions. The research agenda below is designed to close this gap by identifying priority pathways where coordinated, cross-Atlantic work can deliver scalable and interoperable grid architectures.

A cross-Atlantic research program can leverage these assets to develop architectures that are adaptive, scalable, and socially meaningful. The pillars below outline priority collaboration areas that directly address the technical, systemic, and societal challenges documented in Section~\ref{sec_openChallenges}.

\subsection{Collaborative research pillars}\label{subsec_pillars}
The following research pillars define a coordinated U.S.–Danish–EU research agenda for accelerating the deployment of scalable grid architectures and digital flexibility solutions. Because the case studies in Sec.~\ref{subsec_caseExample} are integrative by design, we reference them below as illustrative touchpoints (see Table~\ref{tab:pillars_challenges_cases}) rather than as one-to-one mappings.

\subsubsection*{Pillar 1: Co-design of scalable cyber-physical architectures}
A foundational research need is the co-design of cyber-physical architectures capable of coordinating millions of heterogeneous DERs across diverse ownership models, while ensuring grid reliability, respecting device autonomy, and enabling equitable market participation~\cite{almassalkhi2025pemag}. Flexibility Functions offer a promising AI-based dynamic abstraction that links device behavior with grid- or market-level coordination signals, allowing for both top-down (utility- or DSO-driven) and bottom-up (device- or aggregator-driven) architectures. In addition, FFs can be integrated into laminar architectures to enable scalable coordination across different grid and aggregation layers. 

{\color{black}The case studies in Sec.~\ref{subsec_caseExample} motivate this pillar from complementary angles. Danish Smart Energy OS deployments demonstrate how FFs, hierarchical control layers, and validated digital twins can orchestrate flexible assets from the grid edge to specialized aggregator portfolios. New York’s Grid of the Future proceedings, in turn, highlight the corresponding grid-aware requirements for scaling such coordination in U.S. settings: laminar DSO-Aggregator interfaces and enhanced grid visibility, which enables utilities to compute operating envelopes to ensure grid-aware DER dispatch. A joint research program could therefore generalize these architectural patterns, establish observability/controllability/stability conditions, and develop reference designs that remain valid across heterogeneous devices, ownership models, and regulatory contexts.}

Research thrusts include developing performance certificates for top-down and bottom-up coordination and observability guarantees under partial information (with fewer sensors). Additionally, requirements for large-scale FF-based control under uncertainty, and generative AI-based methodologies for automating model validation and maximizing utilization of both grid and DER assets.

\subsubsection*{Pillar 2: Joint testbeds and digital twins}
Joint U.S.–EU testbeds and validated digital twins are critical to stress-test architectures under scale, heterogeneity, uncertainty, market mechanisms, and device-level algorithms~\cite{ec2024ceeds, cordis2024aieffect}. These testbeds should support end-to-end architectural demonstrations under realistic feeder, customer, and market conditions, including the grid-aware requirements highlighted in Example~1 (visibility, topology/model validation, and operational envelopes) and the field-validated coordination patterns demonstrated in Examples~2-3 (e.g., hierarchical FF-based control).

Numerous U.S.-based laboratories provide distribution- and inverter-level experimentation environments, while Center Denmark provides high-resolution, cross-sector data infrastructures, real-time simulators, and field-deployed pilots (e.g., the summer house demonstrations in Example~2). Interconnecting these assets enables reproducible benchmarking of control schemes, sector-coupled flexibility (electricity-heat-transport), cyber-resilience strategies, and grid-aware activation across diverse operating environments.

Research priorities include methods for developing and cross-validating digital twins across climates and grid topologies (including AI-enabled approaches), benchmark scenarios and data products for cross-comparison, and formalized cyber-physical attack and characterization of the effects of communication delays using federated testbeds.

\subsubsection*{Pillar 3: Standards and interoperability mechanisms}
Interoperability remains the cornerstone of scalable flexibility~\cite{gopstein2021nist1108r4,claes2024mimsplus}. A collaborative research effort must align and reconcile standards across the Atlantic, such as IEEE~1547, IEC~61850, and OpenADR-with emerging European frameworks, including OASC Minimum Interoperability Mechanisms (MIMs) and sectoral EU Data Spaces~\cite{OASC_MIMS,UN_MIMS}. Flexibility Functions can serve as a core MIM that links static market constructs with dynamic device physics, providing a shared technical interface for aggregators, energy communities, DSOs, transmission system operators (TSOs), and market operators.

Research priorities include validating DER coordination across diverse assets and jurisdictions, developing interoperability profiles that link device functionality to communication pathways (including utility-aggregator exchanges needed for operating envelopes and grid-aware dispatch), and designing certification frameworks that ensure predictable, cyber-secure performance at scale. Harmonized testing procedures and compliance criteria developed jointly by U.S. labs and EU standardization bodies would reduce fragmentation and accelerate adoption.

Key tasks include defining minimal interoperability profiles, mapping FFs onto IEEE~1547/IEC~61850/OpenADR pathways, and developing certification test suites that guarantee predictable behavior across large DER populations.

\subsubsection*{Pillar 4: Policy and market design exchange}
Policy and market rules determine which flexibility services can be procured and under what conditions~\cite{eu2024directive1711,eldridge2022_osti1968796}. A cross-Atlantic program should examine structural differences, including U.S. nodal pricing versus EU zonal markets, state-level retail regulation versus national frameworks, and differing interpretations of DSO roles, and how these differences shape feasible coordination architectures and data/telemetry requirements.

Example~1 (New York Grid of the Future) highlights that architecture and governance choices are inseparable: Total-TSO, Hybrid-DSO, and Total-DSO pathways imply different data rights, roles and responsibilities, and coordination protocols. Denmark and broader EU experience with multi-energy integration and evolving flexibility products offers complementary lessons for aligning procurement mechanisms with intertemporal and stochastic flexibility. Interesting research questions include 
(i) TSO-DSO coordination, 
(ii) transparent and equitable aggregator access rules, and 
(iii) procurement and settlement mechanisms that can represent dynamic, time-coupled flexibility without compromising local reliability~\cite{potomaceconomics2025isone2024assessment}.

\subsubsection*{Pillar 5: Workforce and innovation ecosystems}

Scaling flexibility requires both a skilled workforce and robust innovation ecosystems~\cite{doe2024gridmodernization, ec2025dataactexplained}. Implementing grid-aware coordination (e.g., via DSSE, validated models, operating envelopes, interoperability compliance, and cyber-secure DER on-boarding) depends on institutional capacity across utilities/DSOs, aggregators, vendors, regulators, and installers. Joint training programs, student research exchanges, and industrial partnerships can accelerate capacity building and help translate architectures from pilots into durable operations.

Innovation ecosystems such as GreenLab Skive, U.S. national laboratory user facilities, and state-level DER initiatives provide complementary models for de-risking new technologies. Aligned curricula, shared teaching modules, and industry participation in joint testbeds will help ensure that advances in cyber-physical control and interoperability translate into deployable, scalable solutions.

Next, we highlight the role of governance models in shaping roles and responsibilities of different actors in any architecture. Then, we propose the Smart Energy Operating System (SE-OS) as an example of a scalable and adaptive framework that unifies the different pillars. We then conclude this section with three case studies(Sec.~\ref{subsec_caseExample}) that illustrate how the different pillars manifest in energy policy and deployments, from statewide grid planning in New York to device-level coordination and market integration in Denmark, each demonstrating different aspects of scalable architecture design.

\subsection{Governance models for distributed energy resource coordination}\label{subsec_governance}

A scalable architecture for DER coordination must accommodate different ownership structures and regulatory models. 
Emerging experiments across the U.S. and Europe illustrate how different DSO and aggregator roles, responsibilities, and models shape the feasible space for digital flexibility architectures. These issues are central themes in grid-architecture frameworks developed over the past ten years~\cite{TaftFong2021GridArchitecture,GWAC2024Practical} and in analyses of emerging distribution system roles~\cite{DeMartini2015Distribution}. Understanding these models is, therefore, valuable for a cross-Atlantic research agenda aimed at interoperable and scalable DER coordination.

Experiments across the U.S. and Europe reflect the diversity of future DSO roles. For example, in Vermont, Green Mountain Power (DSO) is adopting a \emph{Total DSO} architecture by directly owning, leasing, and orchestrating DERs (sometimes through third-party aggregators) to provide wholesale market services, active load shaping, and local resilience (including full-feeder islanding and behind-the-meter battery programs). Other U.S. states are adopting \emph{Hybrid DSO} architectures through aggregator-driven frameworks in which utilities focus on maintaining grid reliability, while aggregators manage DER participation in energy markets and resilience services. These distinctions mirror architectural separations emphasized in the literature, particularly the delineation between coordination, control, and responsibility layers in modern distribution systems~\cite{DeMartini2015Distribution,TaftFong2021GridArchitecture}.

These models illustrate two competing visions of the future: a utility-centric Total DSO coordination model and an aggregator-centric Hybrid DSO model. Regulating these models to ensure equity, reliability, and efficient market access remains an open research question. Concepts, such as, distribution-level locational marginal pricing (DLMPs)~\cite{Caramanis2016DLMP}, aggregator participation rules, and automated DER controls offer possible pathways, but their impacts on grid stability and consumer welfare need to be evaluated rigorously through joint testbeds and digital twins~\cite{GWAC2024Practical}.

\subsection{Proposed framework: The Smart Energy Operating System} \label{subsec_seos_method}
{\color{black} 
The governance models in Sec.~\ref{subsec_governance} determine who coordinates DERs and which information must cross organizational boundaries, while the technical pillars in Sec.~\ref{subsec_pillars} identify the cyber--physical capabilities needed to make that coordination scalable. Together, they motivate the \emph{Smart Energy Operating System} (SE-OS): a reference operational stack and methodological framework that unifies data, models, optimization, control, and interoperability for multi-level demand-side flexibility coordination~\cite{tohidi2025seos}. SE-OS specifies (i) actor roles and responsibilities across grid operators, aggregators, and devices, (ii) the minimum interoperability mechanisms (MIMs) and exchange pathways required for reliable coordination, and (iii) a workflow that translates system objectives and constraints into grid-admissible device-level actions with measurable performance.
}


Architectural approaches to distributed flexibility require coherent coordination across devices, aggregators, DSOs and TSOs and in analyses of emerging DSO roles~\cite{TaftFong2021GridArchitecture,GWAC2024Practical,DeMartini2015Distribution}. The principles of forecasting, control, optimization, and interoperability described above converge in the concept of the \emph{Smart Energy Operating System} or Smart Energy OS (SE-OS). The SE-OS provides a coherent framework for developing, implementing, and testing multi-level DSF coordination schemes across data, models, optimization, control, and communication layers~\cite{madsen2015a, madina2019a, de2018a, tohidi2025c}. Its central purpose is to enable the hierarchical operation of flexible energy systems across all spatial and temporal scales, from individual appliances to buildings, districts, cities, and national systems.

A consistent requirement for these hierarchies is that actions taken at the transmission, distribution, and device levels remain coherent. This reflects the \emph{laminar} architectural principle that each grid layer should maintain clear roles, boundaries, and well-defined interfaces, thereby revealing hidden couplings and preventing tier bypassing~\cite{TaftFong2021GridArchitecture,GWAC2024Practical}. For example, the hierarchical wind forecasting method in \cite{hansen2023} produces significant improvements in accuracy while ensuring that forecasts used by the TSO and DSOs are mutually consistent. Similarly,~\cite{aschenbruck2023a} formulates interfaces between transmission, distribution, and microgrid layers and shows how flexibility originating in microgrids, such as community batteries, can be activated at higher aggregation levels. A related modeling framework in \cite{tso-dso-ff} demonstrates how distributed flexibility can be integrated across multiple hierarchy levels using a unified mathematical representation.

A defining feature of the Smart Energy OS is the use of Flexibility Functions as a \emph{Minimum Interoperability Mechanism} (MIM). The FF provides a minimal and sufficient interface that links device- and building-level physics to upper-level control or market signals. By sharing only the information necessary for coordination, FFs preserve the hierarchical boundaries described above and enable interoperability without requiring disclosure of detailed models or operational data. The same laminar interface applies consistently at all aggregation levels, including appliances, buildings, districts, cities, and larger regions, which provides a unifying language for flexibility across the entire hierarchy.

The Smart Energy OS operates on a {\color{black} hierarchy of forecasts and models arranged in a coherent and consistent hierarchy from the edge to the cloud (see, e.g., \cite{nystrup2020temporal})}. Data and control typically remain close to the device at the edge (e.g., in mobile phones, smart home systems, or embedded controllers), while fog and cloud layers provide aggregation, coordination, and forecasting services. Computation is distributed across these layers, and the architecture is explicitly designed to respect privacy, transparency, and fairness. One-way broadcast price signals serve as a privacy-preserving control mechanism, allowing buildings and devices to self-dispatch based on local preferences without exposing sensitive data (upstream).

Data spaces and federated information-sharing are the core of Smart Energy OS, {\color{black} and concepts are further developed in the new European Energy Data Space project (INSIEME, 2025-2028) \cite{insieme2025}}. Denmark’s national digitalization initiative {\color{black}{FDP 2024-2027, \cite{fdp2025}}}, for example, integrates electricity and heat data to support cross-sector coordination. These data spaces ensure coherent spatial and temporal hierarchies for system-level forecasting, such as the multi-level wind generation forecasts illustrated in \cite{sorensen2023a}. {\color{black} The Flexibility Function is often used as a dynamic mapping between price signals and load signals. However, the FF concepts can be used to forecast other features such as the voltage in a DSO grid}. Although comparable nationwide data infrastructures do not yet exist in the United States, emerging state-level initiatives (e.g., New York’s DER inventory and hosting capacity data) are moving in a similar direction.

\begin{figure}[t]
    \centering
    \includegraphics[width=0.9\linewidth]{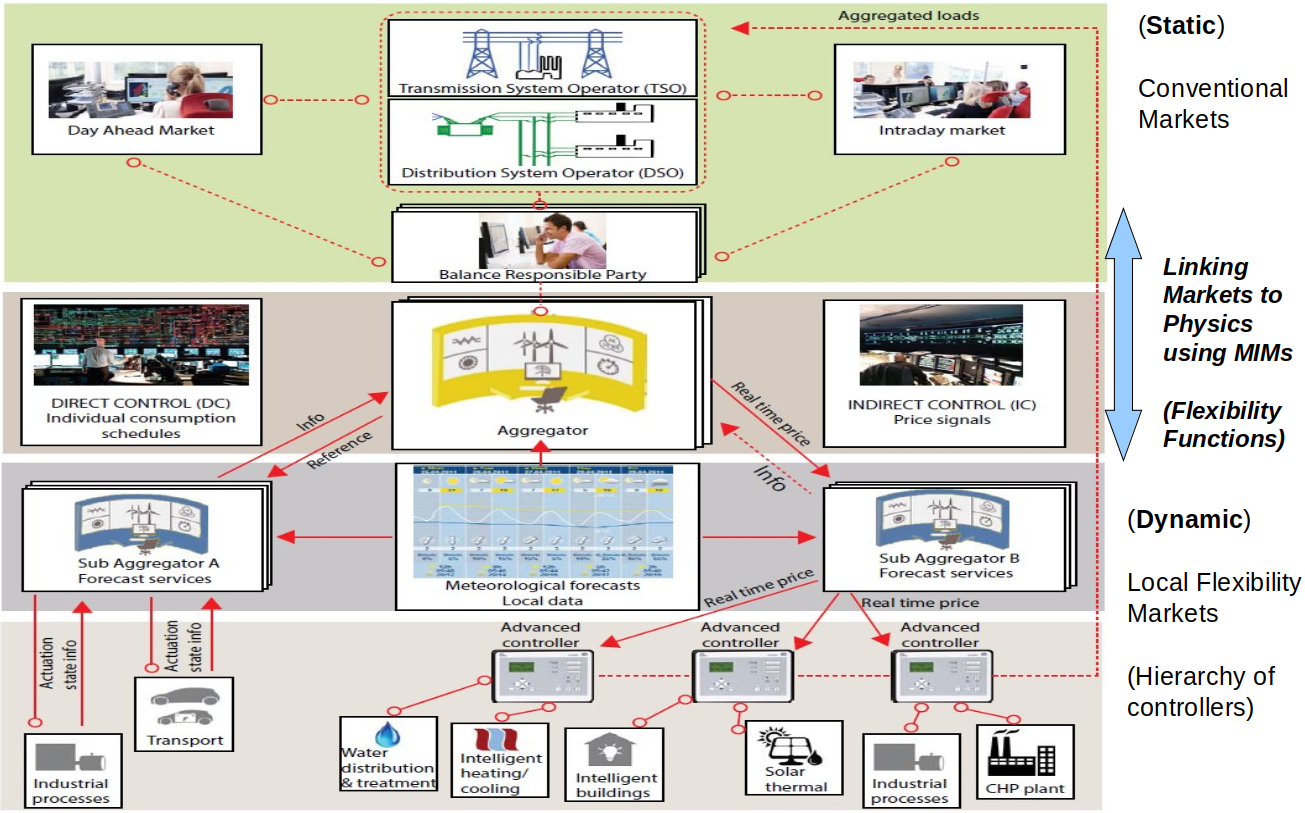}
    \caption{Schematic of the Smart Energy Operating System (SE-OS), showing hierarchical coordination across various layers and the interfaces that connect device-level flexibility with distribution and system-level operations. {\color{black} Notice that the Flexibility Function is used for linking markets to the physics at lower-levels. The Flexibility Function can be considered as a data model within the Minimum Interoperability Mechanisms (MIMs).}}
    \label{fig:smart_energy_os}
\end{figure}

{\color{black}Flexibility Functions provide a key element in providing an operational link between conventional markets and controllers for providing grid and balancing services (see also Figure \ref{fig:smart_energy_os}). The FF concepts can also be used for demand-response solutions in district heating \cite{mokhtari2025a,mokhtari2025b}. Finally,} the FF also provides a natural bridge for sector coupling. Buildings equipped with both district heating and electric heat pumps, for instance, can automatically switch between energy carriers depending on price or carbon intensity, enabling cost-effective decarbonization and supporting grid balancing. Similarly, hybrid energy systems involving thermal storage, EV charging, or power-to-X resources can participate in coordinated flexibility mechanisms without the need for bespoke integration at each site.

The hierarchical Smart Energy OS architecture is illustrated in Fig.~\ref{fig:smart_energy_os}. At higher layers, conventional markets continue to operate using established mechanisms such as day-ahead and intra-day auctions. At lower layers, the OS uses both direct (setpoints) and indirect (price-based) control signals to activate DER flexibility. In practice, most building-related demand response in Denmark relies on indirect price-based control, which reduces communication requirements, preserves end-user autonomy, and aligns with the laminar, responsibility-preserving structure described earlier.

Several European and international projects have validated Smart Energy OS concepts. The \textsc{ebalance+} project~\cite{mohsen2023}, Flexible Energy Denmark~\cite{blomgren2023a}, and the EU H2020 SmartNet project~\cite{madina2019a} demonstrate coordinated activation of flexibility through hierarchies of controllers across residential, commercial, and industrial sectors. These demonstrations show how aggregated flexibility information from Home Energy Management Systems (HEMS) and Home Management Information Systems (HMIS) can be used to provide grid services, optimize energy efficiency, and support local energy communities.

Overall, the Smart Energy OS complements the research pillars by providing:
\begin{enumerate}[label=(\roman*)]
    \item a unifying cyber-physical architecture for scalable coordination (Pillar~1),
    \item a natural structure for joint testbeds and federated digital twins (Pillar~2),
    \item a concrete embodiment of interoperability mechanisms, including MIMs and FFs (Pillar~3),
    \item a pathway for integrating dynamic and stochastic market signals with real-time control (Pillar~4), and
    \item an applied context for training the next-generation digital energy workforce {\color{black}towards a future self-operation grid \cite{tohidi2025a}} (Pillar~5).
\end{enumerate}

These capabilities highlight how a laminar architecture can coherently unify device-level actuation, distribution-level observability, and system-level coordination. To illustrate how these architectural principles manifest in practice across different regulatory and technological contexts, we now present three demonstrations from the United States and Denmark.

\subsection{Real-world case studies}\label{subsec_caseExample}

To ground the architectural concepts discussed above, this section presents three real-world demonstrations that illustrate how distributed flexibility can be integrated into system operations. The first example examines New York State’s \emph{Grid of the Future} proceedings, which evaluates how DERs can be coordinated at scale within U.S. regulatory and market structures to enhance affordability, reliability, and planning efficiency~\cite{GOTF_Phase1_Vol2_2025,GOTF_Phase1_Vol3_2025}. The second and third examples highlight field deployments in Denmark enabled by the Smart Energy OS. Example~2 demonstrates how coordinated flexibility from summer houses with heated pools can support both TSO and DSO objectives. Example~3 illustrates how specialized aggregators can leverage Flexibility Functions to provide market services while reducing operational costs for end users.
{\color{black}
Table~\ref{tab:pillars_challenges_cases} provides a high-level mapping between the research pillars in Sec.~\ref{subsec_pillars}, the challenge categories from Sec.~\ref{sec_openChallenges}, and the illustrative case studies (Examples~1--3) described next. Notably, because the case studies are integrative by design, the mapping is non-exclusive and reflects the \emph{primary} emphasis of each example.
}

\begin{table}[t]
{\color{black}
\caption{Highlighting \emph{primary} links between challenge focus (Sec.~\ref{sec_openChallenges}) \& pillars (Sec. \ref{subsec_pillars}) and the case-study examples (Sec.~\ref{subsec_caseExample}).}
\label{tab:pillars_challenges_cases}
\scriptsize
\setlength{\tabcolsep}{3pt}
\renewcommand{\arraystretch}{1.0}
\centering
\begin{tabularx}{\columnwidth}{@{}p{0.75cm}Yp{2.0cm}@{}}
\toprule
\textbf{Pillar} & \textbf{Challenge focus (keywords)} & \textbf{Case study} \\
\midrule
P1 & Scalable control co-design; stability/synchronization; observability needs & Examples 1--3 \\
P2 & Digital twins/testbeds; validation; M\&V; DSSE/telemetry realism & Example~2 \\
P3 & Interoperability profiles; FF/MIM as interface; secure data exchange & Examples 2--3 \\
P4 & Market \& governance heterogeneity; TSO--DSO coordination; procurement rules & Example 1 \\
P5 & Adoption + workforce; deployment friction; equity/privacy constraints & Examples 1--2 \\
\bottomrule
\end{tabularx}
}
\vspace{2pt}
{\footnotesize \textit{Legend:} P=Pillar  (Sec.~\ref{subsec_pillars}); Case study examples (Sec.~\ref{subsec_caseExample}).}
\end{table}

\subsubsection{Example~1 (Grid of the Future): Scalable distributed energy resource integration in New York}

New York’s \emph{Grid of the Future} (GOTF) initiative provides one of the most comprehensive U.S. efforts to assess how distributed flexibility can support decarbonization, electrification, and reliability at scale. Phase~1 of the study, led by The Brattle Group and NYSERDA, identified more than \emph{8~GW of cost-effective, hourly flexibility potential} by 2040~\cite{GOTF_Phase1_Vol3_2025}. This came primarily from  EV charging, HVAC, battery storage, and commercial and industrial demand response, illustrating both the scale and diversity of DERs available for system balancing~\cite{GOTF_Phase1_Vol3_2025}. Brattle estimated that coordinated use of this flexibility could deliver \emph{approximately \$3~billion per year} in system value, driven largely  by avoided generation and distribution capacity costs, underscoring the importance of \emph{temporal coordination} for shifting demand away from peak periods. About 80\% of savings would be returned to participants, representing a potential affordability boost, if equitably deployed.

At the same time, Phase~1 highlighted that flexibility benefits are highly \emph{location-dependent}: nearly half of New York’s substations may face thermal or voltage constraints under future electrification scenarios. As a result, DERs cannot be coordinated solely through wholesale market signals; scalable integration requires \emph{distribution-aware architectures} that account for feeder-level limits and localized congestion. Brattle further noted that low interconnection rates, limited DER visibility, and inconsistent modeling practices hinder current flexibility deployment, reinforcing an overarching architectural insight: \emph{good system models beget good control}. Collectively, these findings motivate the need for interoperable data pathways, standardized operating envelopes, and cyber–physical coordination mechanisms capable of managing DER flexibility at scale. This observation directly motivates the architectural approaches in Section~\ref{sec:hierarchical_control} and aligns with recent analyses of cyber–physical challenges with VPPs~\cite{almassalkhi2025pemag}.

\begin{figure}[t]
    \centering
    \begin{subfigure}[t]{0.95\linewidth}
        \centering
        \captionsetup{width=\linewidth}
        \includegraphics[width=\linewidth]{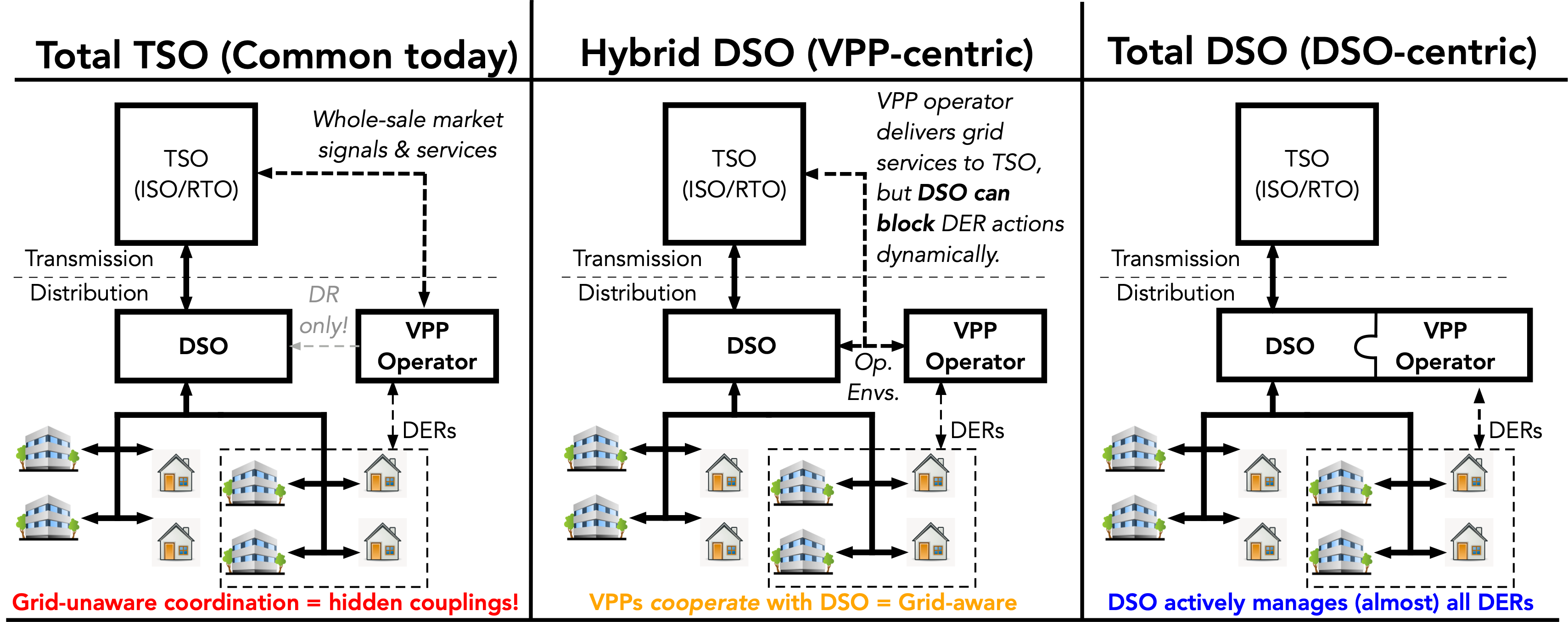}
        \caption{Architectural models under consideration in New York’s Grid of the Future proceedings. These models reflect varying degrees of utility and aggregator responsibility for DER dispatch, system visibility, and grid-aware coordination.}
        \label{fig:GOTF_models}
    \end{subfigure}

    \vspace{1em}

    \begin{subfigure}[t]{0.5\linewidth}
        \centering
        \captionsetup{width=\linewidth}
        \includegraphics[width=\linewidth]{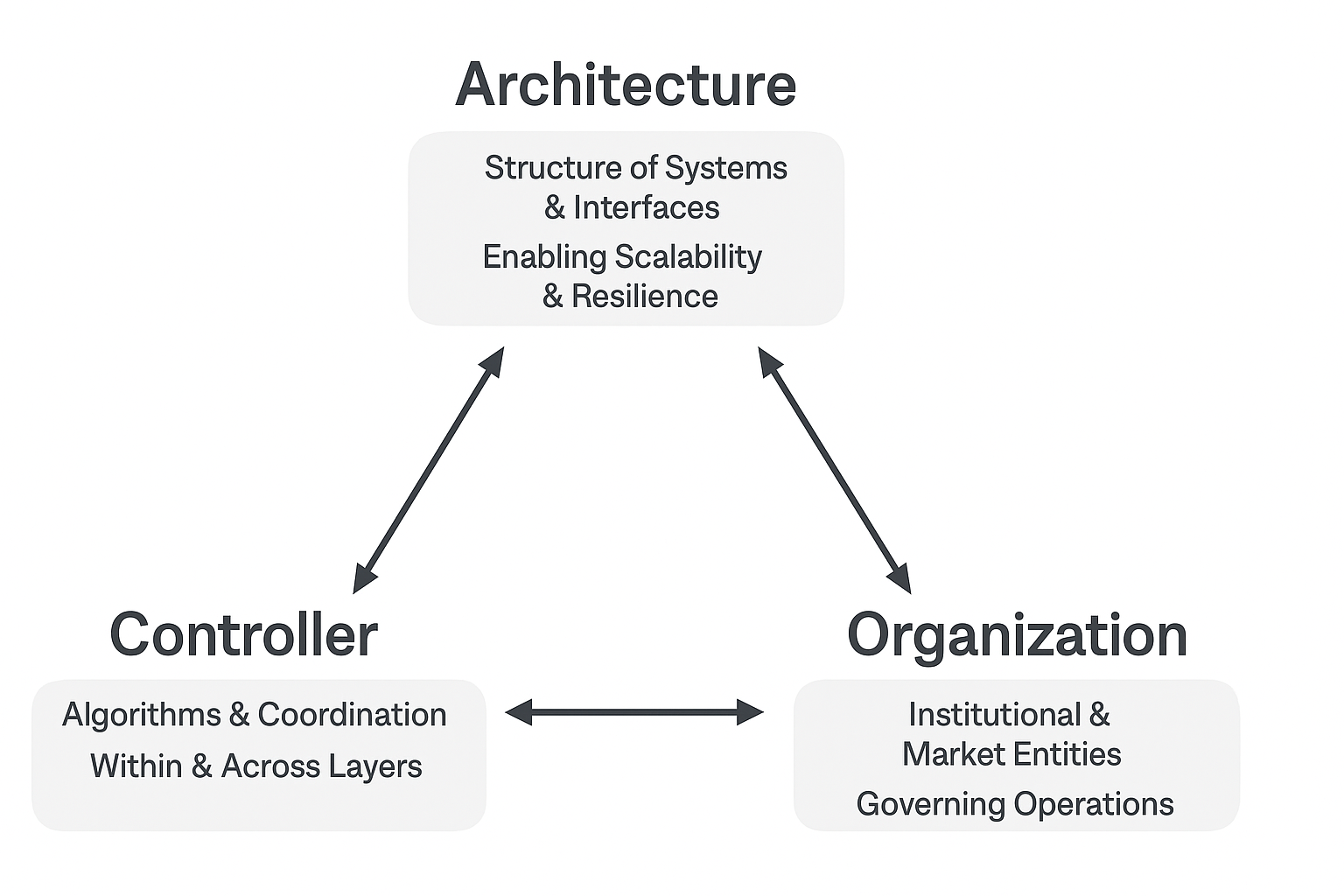}
        \caption{Relationship between Architecture, Controller, and Organization. 
        Scalable DER integration requires alignment between system structure (architecture),
        coordination mechanisms (controller), and institutional roles (organization).}
        \label{fig:GOTF_arch_triangle}
    \end{subfigure}
    \caption{Grids of the future will require systematic design for interfacing organizations, their data, and control layers within a scalable and adaptive system of systems.}
\end{figure}

Phase~2 of the GOTF initiative reviewed utility Distribution System Implementation Plans (DSIPs) and underscored a central architectural requirement for scalable flexibility: \emph{distribution networks must be observable, interoperable, and data-accessible}.  
The findings emphasized that real-time visibility, including telemetry, DSSE, and validated topology and load models, is essential for grid awareness and for unlocking DER flexibility without compromising reliability.

New York is currently advancing Phases 3 and 4 to categorize future capabilities unlocked by DERs and to define salient grid architectures. Specifically, New York is considering different architectural pathways for large-scale DER integration. First, a \emph{Total TSO} or \emph{market-only} model (Fig.~\ref{fig:GOTF_models}, left) allows DERs 
to respond directly to wholesale market signals with little or no distribution-level coordination. While simple to implement, this approach becomes increasingly problematic at high DER penetration, since the DSO lacks the visibility or authority needed to prevent congestion, voltage violations, or tier bypassing.  

Second, a \emph{Hybrid DSO} model (Fig.~\ref{fig:GOTF_models}, center) preserves aggregator (VPP) autonomy but actively constrains their dispatch through utility-defined operating envelopes or dynamic hosting capacities that are based on actual/estimated grid conditions. This model balances innovation with reliability and is popular in practice in the U.S. today, because it can provide (some) grid-awareness while maintaining competitive aggregator participation. Open questions still remain about competitiveness of participation under high DER penetration when grid reliability blocks significant aggregator dispatches and it is challenging to unlock critical reliability benefits within a (competitive) market structure.

Third, a \emph{Total DSO} model (Fig.~\ref{fig:GOTF_models}, right) gives the utility full responsibility for DER siting, deployment, and orchestration for both reliability and economic services. This model requires strong DSO visibility, integration of distributed energy resource management systems (DERMS) and advanced distribution management systems (ADMS), and feeder-level constraint management, and is well-suited to high-DER feeders where continuous operational oversight is needed. Furthermore, with DSO incorporating DSF into planning and operations, reliability benefits can be realized directly, increasing the value from DERs.

Across these architectural options, a consistent conclusion from GOTF is that scalable DER integration requires clear aggregator–utility interfaces, transparent operating limits, and real-time grid visibility to avoid hidden couplings and grid-unaware dispatch. These issues can be interpreted through the lens of architectural alignment shown in Fig.~\ref{fig:GOTF_arch_triangle}, which illustrates how system performance depends on coherence between: (i) the structural design of the grid and its interfaces (\emph{architecture}), (ii) coordination and activation mechanisms (\emph{controller}), and (iii) institutional and market roles  (\emph{organization}). GOTF demonstrates that misalignment among these dimensions, e.g., aggregators controlling DERs without grid visibility, or utilities lacking data needed to enforce constraints, prevents scalability even when technical flexibility exists.

Overall, New York’s GOTF effort demonstrates a central theme of this research agenda: \emph{scalable architectures must unify distribution constraints, device-level dynamics, and market participation through transparent data pathways and coordinated cyber–physical control}. As DER penetration grows, architectural considerations, including grid visibility, laminar role separation, and standardized interoperability mechanisms, increasingly determine how flexibility can be integrated safely and cost-effectively into system operations.






The GOTF efforts in New York underscore that scalable flexibility ultimately depends on architectures that couple device-level actuation with distribution-level observability and clear institutional interfaces. The following two examples from Denmark illustrate how these same architectural principles are implemented in practice through the Smart Energy OS: first at the device and home level, and then at the aggregator level, where coordinated flexibility can provide balancing services and substantial cost reductions.

\subsubsection{Example 2 (Smart Energy Operating System): Energy flexibility from Danish summer houses}


\begin{figure}[t]
\centering
\begin{subfigure}[t]{0.38\linewidth}
    \centering
    \includegraphics[width=\linewidth]{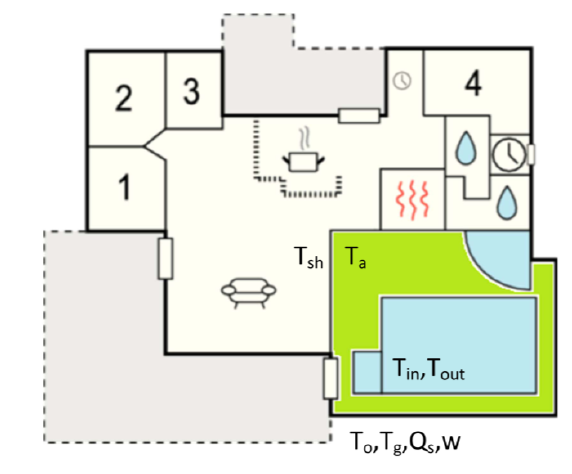}
    \caption{Floor plan with heated pool, water heater, and key sensors/actuators with electrothermal parameters $T,Q,w$.}
    \label{fig:ground_plan}
\end{subfigure}
\hfill
\begin{subfigure}[t]{0.60\linewidth}
    \centering
    \includegraphics[width=\linewidth]{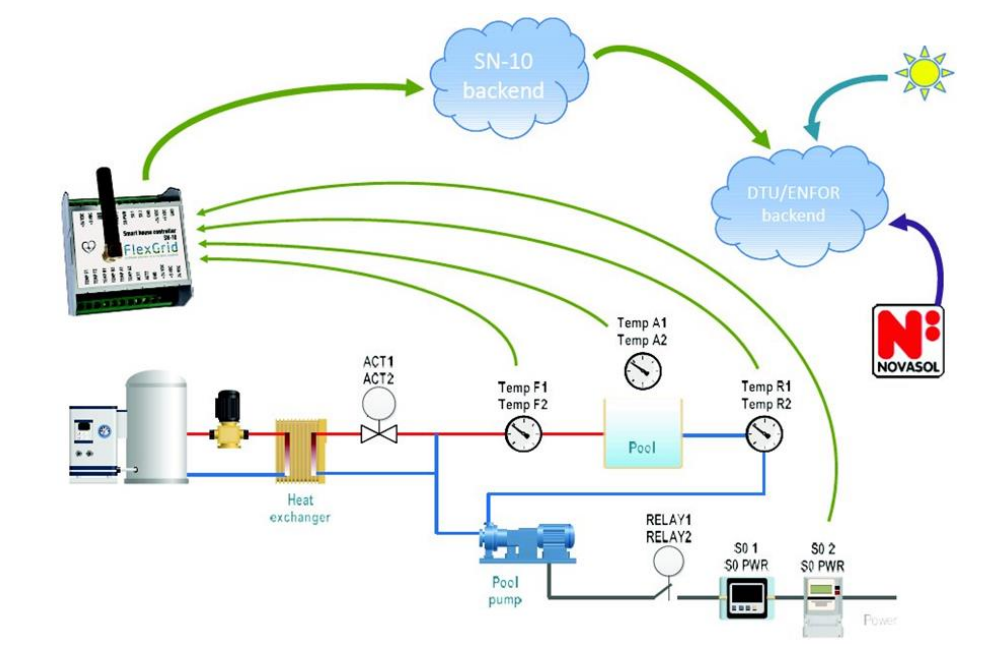}
    \caption{Cyber-physical implementation with FlexGrid controller, data aggregation, and pool system components.}
    \label{fig:data_gathering}
\end{subfigure}
\caption{Physical and cyber-physical components of the summer house pilot demonstrating FF-based control in the Smart Energy OS.}
\label{fig:summerhouse_overview}
\end{figure}

In 2022, more than 63\% of Denmark’s electricity consumption was supplied by intermittent wind and solar generation~\cite{Energinet2025Miljoredegorelse}. This variability introduces balancing challenges that motivate the need for DSF that can shift consumption without impacting comfort. Danish summer houses with indoor swimming pools provide a particularly promising opportunity: across Denmark, more than 900 such homes exhibit substantial, predictable, and highly deferrable electric demand, largely through pool heating and humidity control. Their large thermal mass allows heating to be shifted over multiple hours with minimal comfort impacts, making them well-suited for ancillary and balancing services.

NOVASOL, Denmark’s largest organizer of summer house rentals, participated in the project due to its interest in reducing electricity costs for owners and renters. This industry involvement reflects a broader, customer-facing value proposition for flexibility that aligns with grid needs.

The pilot project described here is part of a number of R\&D projects, including SmartNet, \textsc{ebalance}+, CITIES, and FED initiatives, and tests how the Smart Energy OS framework can activate this flexibility for both DSO- and TSO-level services. Energinet (TSO) and N1 (DSO) participated to demonstrate price-based indirect control using FFs. A representative set of 30 homes in Blokhus and Blåvand was selected for initial field trials: each DSO area formed its own cluster, while the TSO viewed all 30 homes as a single aggregated resource. This represents a rare end-to-end demonstration of FF-based hierarchical coordination deployed on actual households.

\begin{figure}[t]
\centering
\includegraphics[width=0.7\linewidth]{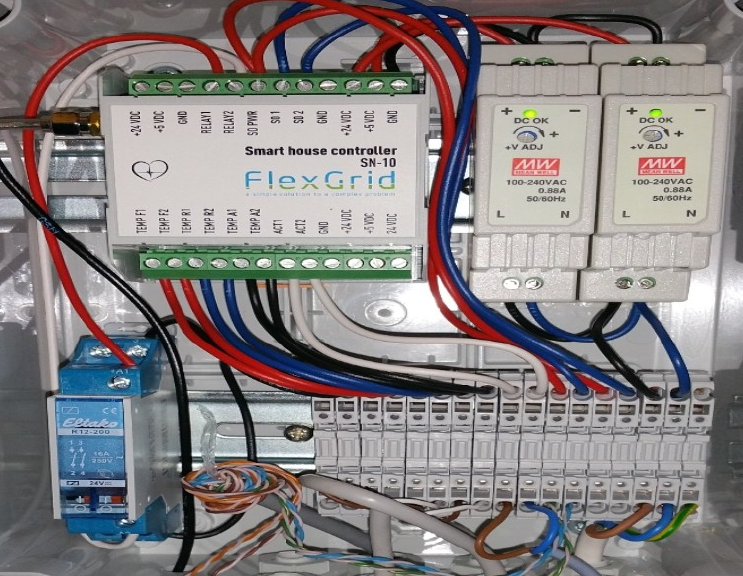}
\caption{Cellular-connected SN-10 gateway used for local sensing and actuation.}
\label{fig:SN_10_gateway}
\end{figure}

An overview of the physical and cyber-physical components of the pilot project are shown in Fig.~\ref{fig:summerhouse_overview}.
Each participating home was equipped with an SN-10 gateway (see Fig.~\ref{fig:SN_10_gateway}) that transmitted temperature, humidity, and pool state data to a cloud-based Smart Energy Hub. ENFOR provided forecasting and optimization, while the Smart Energy OS executed local control based on dynamic price broadcasts derived from FF-based models and operational constraints as in Fig.~\ref{fig:setup_aggregator}.

\begin{figure*}[t]
\centering
\includegraphics[width=\linewidth]{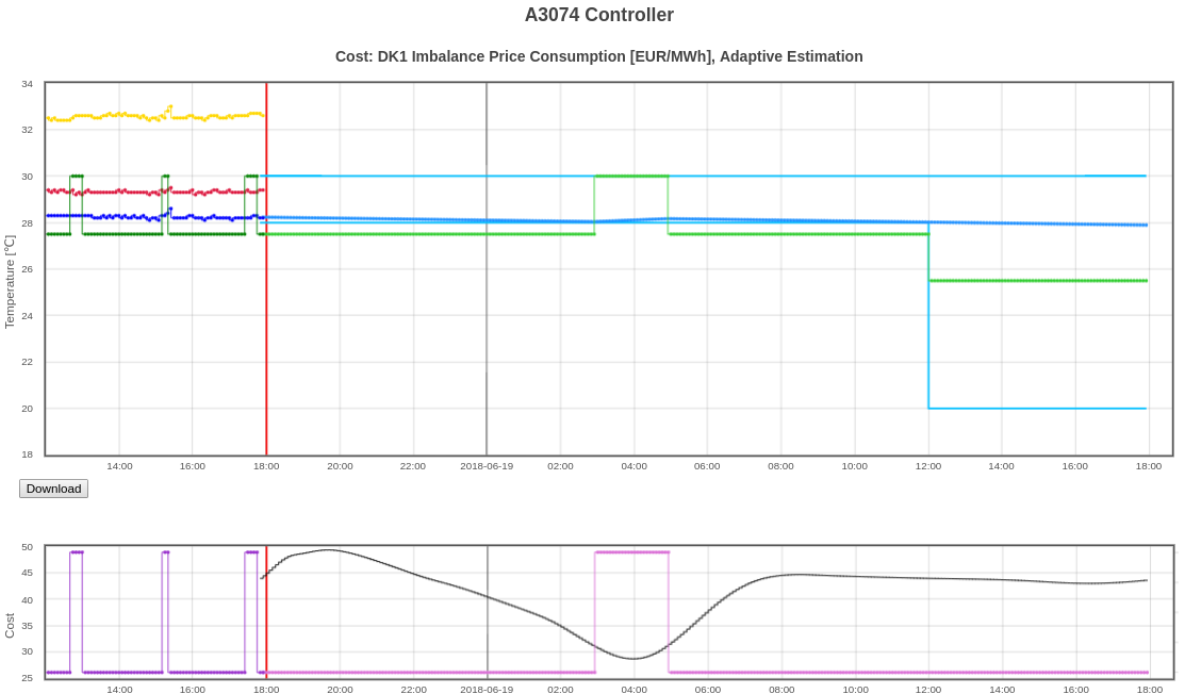}
\caption{Forecasting and price-based control for an individual house (A3074), using DK1 imbalance prices. {\color{black}The red vertical line represents the current time with data to the left being historical data while data to the right is predicted response. The top plot shows the forecasts of the temperature given the optimal price-based control depicted in the bottom plot. The heating is optimized for when  the imbalance prices (black line in bottom plot) are lowest.}}
\label{fig:forecasting_control}
\end{figure*}

Price-based indirect control was used: the Smart Energy OS computed a dynamic price signal based on forecasts of grid load, wholesale and imbalance prices, weather conditions, and rental schedules. This signal was then broadcast to the fleet of summer houses, each of which self-dispatched according to its dynamics-aware, FF-based thermal model and its local comfort constraints.

Owners selected their preferred operational mode. Most prioritized electricity cost savings; others occasionally chose a ``green operation'' mode that minimized carbon intensity. Figure~\ref{fig:forecasting_control} shows an example from house A3074, where heating is optimally scheduled during low-price nighttime periods. The thermal inertia of the pool allows temperature to drift gradually back toward its lower limit during the day, maintaining comfort while reducing energy cost.

Across the fleet, reported electricity cost and emissions reductions ranged from 20–35\%. 
This example illustrates several core architectural themes of this paper:
\begin{enumerate}[label=(\roman*)]
    \item \emph{a lightweight, standardized interface} (the Flexibility Function) that enables aggregators and operators to coordinate without intrusive data exchange;
    \item \emph{hierarchical control architectures} that support both TSO- and DSO-level services through coherent price-based indirect control;
    \item \emph{low-cost edge hardware and data-driven digital twins} that make individual DERs predictable, observable, and controllable; and
    \item \emph{simple broadcast signals} that unlock meaningful, real-time system flexibility with minimal user involvement.
\end{enumerate}

This pilot demonstrates how Smart Energy OS principles of interoperability, hierarchy, FF-based TSO-DSO coordination, and privacy-preserving control can be deployed in practice to harness device-level flexibility in an operator-friendly manner. 
It also provides a rare, real-world view into the architecture-level requirements that enable DER coordination to potentially scale reliably, including edge-level observability, validated digital twins, and one-way broadcast control.

\subsubsection{Example 3 (Smart Energy Operating System): Specialized aggregators for grid services}

\begin{figure}[t]
\centering
\includegraphics[width=\linewidth]{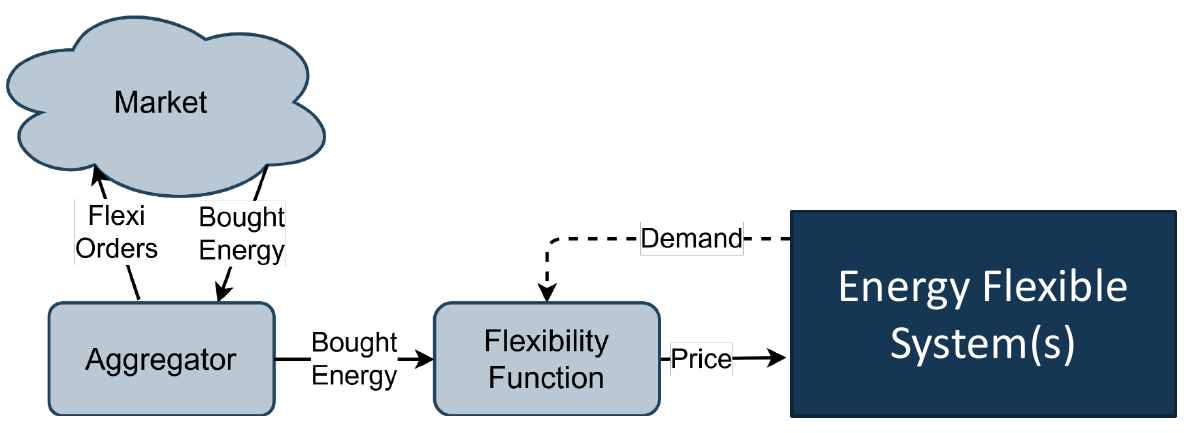}
\caption{Smart Energy OS setup for a specialized aggregator using the Flexibility Function for market participation.}
\label{fig:setup_aggregator}
\end{figure}

The Smart Energy OS also enables a new class of \emph{specialized aggregators} that leverage domain expertise--for example, in summer houses, wastewater treatment plants, or supermarkets--to coordinate flexibility across a portfolio of similar assets. Within the Smart Energy OS hierarchy, these aggregators use the Flexibility Function to (i) optimize procurement in wholesale markets and (ii) translate system-level signals into device-level broadcast prices that elicit a coordinated and grid-aware response.

In Scandinavian electricity markets, one mechanism available to aggregators is the ``Flexi Order,'' defined by an energy amount, a delivery window, and a duration. For example, an aggregator might commit to purchasing 1~MWh in the cheapest 2~hours between 08:00 and 12:00. Flexi Orders can be combined with traditional spot market bids, and the Smart Energy OS uses the FF to determine:  
(i) when to shift consumption within the window, and  
(ii) how to broadcast the appropriate price signal to participating devices.

\begin{figure}[t]
\centering
\includegraphics[width=\linewidth]{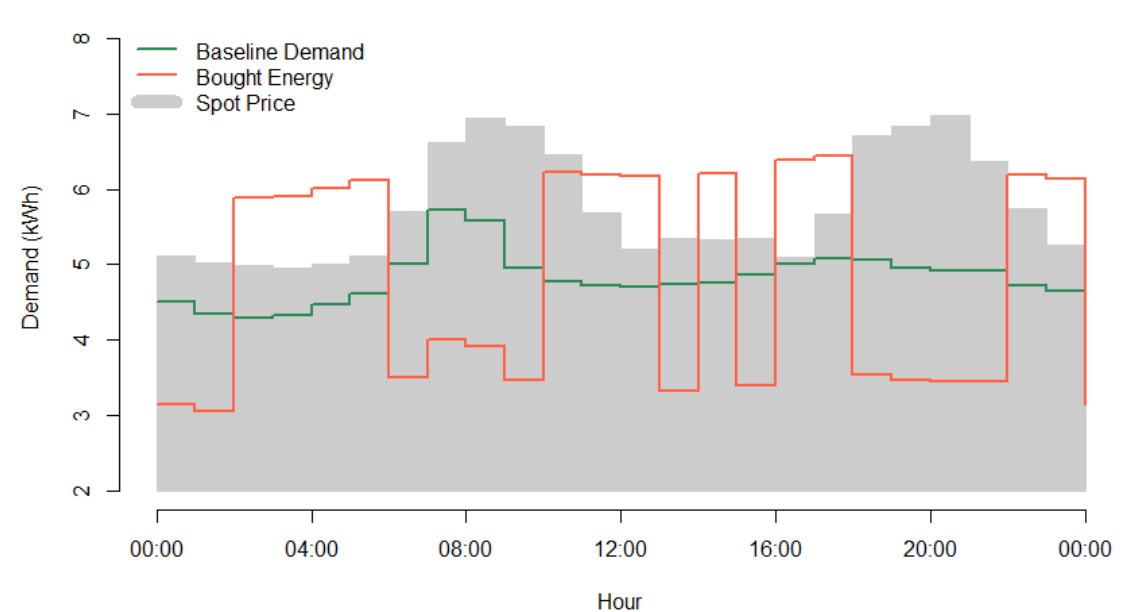}
\caption{Spot price, baseline load, and optimized purchased energy under the Smart Energy OS.}
\label{fig:bought_energy}
\end{figure}

\begin{figure}[t]
\centering
\includegraphics[width=\linewidth]{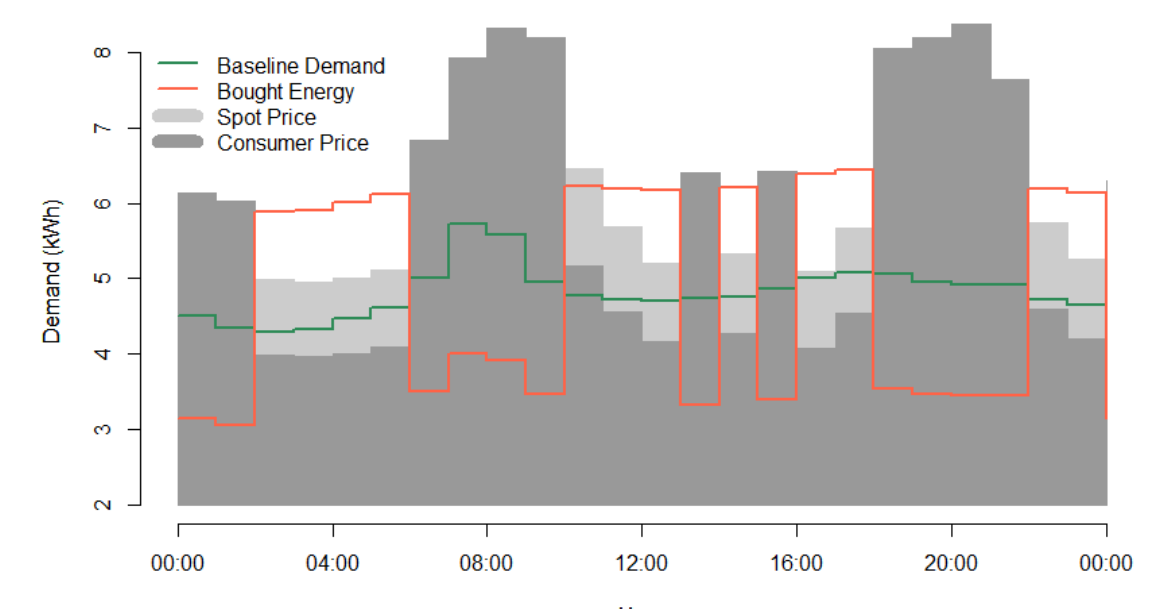}
\caption{Price signal broadcast to end-users for coordinated activation of flexibility.}
\label{fig:consumer_price}
\end{figure}

Figure~\ref{fig:setup_aggregator} illustrates the general setup. The aggregator uses the FF to forecast the dynamic load response of its resource pool and to determine optimal wholesale purchases. Figure~\ref{fig:bought_energy} shows how the aggregator increases consumption when prices are low and reduces it when prices are high. Figure~\ref{fig:consumer_price} then shows the corresponding broadcast price signal sent to participating devices.

This architecture has been applied in Denmark to multiple resource types. For example, \cite{junker2020a} demonstrates cost savings for water tower operations, while \cite{amaral2020a} applies similar principles to water pumping. 
Across these trials, these deployments reveal four recurring architectural features of the Smart Energy OS:  
\begin{enumerate}[label=(\roman*)]
    \item \emph{portfolio-level optimization} driven by wholesale and balancing market signals;
    \item \emph{cluster-level broadcast control} implemented through Flexibility Functions, enabling coherent group-level actuation;
    \item \emph{device autonomy and privacy}, since each asset self-dispatches based on local models and constraints; and
    \item \emph{scalable aggregation}, as new assets can be added to the portfolio without redesigning or recalibrating the overarching control architecture.
\end{enumerate}

This example demonstrates a complete market-to-device, FF-based control loop operating in a real 
balancing market. It shows how standardized Flexibility Functions and minimum-interoperability 
mechanisms enable bottom-up market participation while preserving grid reliability. Taken together, 
the aggregator-driven demonstrations illustrate a central message of this paper: scalable 
flexibility requires architectures that allow TSOs, DSOs, aggregators, and end-users to coordinate 
through transparent and low-friction interfaces, in a manner compatible with both market-based and 
physics-based control needs.

\subsection*{Synthesis and implications}

Collectively, these examples show that scaling DER flexibility is not a device problem but an architectural one. It requires coherent interfaces, hierarchical control, and data pathways that  operate seamlessly across local devices, aggregators, DSOs, and TSOs. New York’s Grid of the Future Proceedings demonstrates the system-level value of flexibility and the importance of distribution grid-aware  coordination. The Danish summer house pilots show how lightweight, device-level abstractions, such as the Flexibility Function, can unlock reliable actuation at the edge of the grid. The specialized aggregator example highlights how these same principles can scale upward to provide market services, balancing support, and substantial energy cost reductions.

Viewed collectively, these case studies demonstrate the complementarity of U.S. and Danish approaches. The U.S. provides regulatory diversity, large-scale deployment environments, and  rapidly evolving aggregator ecosystems, while Denmark offers deep experience with digitalization, sector coupling, and hierarchical control. Their integration underscores the central message of this  research agenda: \textit{realizing the full value of distributed flexibility requires interoperable  cyber–physical architectures that function coherently across devices, aggregators, and system  operators.} This cross-Atlantic perspective sets the stage for the concluding discussion on future  pathways and collaborative opportunities. Together, these demonstrations provide concrete validation  points for the research pillars and highlight the urgency of developing interoperable and scalable architectures using advanced data- and AI-based methodologies in both regions.


\section{Conclusion and Call to Action}
\label{sec_conclusion}
{\color{black}
The clean energy transition is rapidly transforming electricity systems into deeply digital, cyber--physical infrastructures in which distributed flexibility is a core operational resource. This paper’s central finding is that \emph{scalable flexibility is not a device attribute but an architectural property}: it depends on coherent alignment between institutional roles, data pathways, and coordination interfaces across TSO/DSO/aggregator/grid-edge layers.

Achieving this alignment enables New York’s Grid of the Future vision, which has identified \emph{more than 8~GW} of cost-effective, hourly flexibility potential by 2040 and estimated \emph{approximately \$3B/year} in system value from coordinated flexibility (with \emph{about 80\%} of savings returned to participants). In Denmark, scalable flexibility can help integrate more wind and solar affordably as shown in the  Smart Energy OS pilot, where \emph{30} coordinated pool-equipped summer houses reported electricity cost and emissions reductions across the fleet of \emph{20--35\%}. 

These results reinforce the paper’s architectural implications: (a) flexibility value is realized only when activation is \emph{grid-aware} in time and space; (b) predictable actuation at scale requires interoperable interfaces (e.g., FF-style abstractions), validated models/digital twins, and transparent data/telemetry pathways; and (c) TSO-DSO-aggregator coordination must be supported by governance and market mechanisms that translate system objectives into admissible device-level actions.

To accelerate progress towards deployable next-generation grid architectures, future research direction will consider the following:
\begin{enumerate}
  \item \textbf{Joint testbeds and federated pilots:} link DK Smart Energy OS deployments with U.S. utility pilots and R\&D platforms to evaluate FF-style coordination and distribution-aware actuation in U.S. operating conditions.
  \item \textbf{Harmonized interoperability mechanisms:} develop reference interoperability profiles that map coordination abstractions onto existing standards like IEEE~1547, IEC~61850, and OpenADR, and emerging European Minimum Interoperability Mechanisms.
  \item \textbf{Coordinated policy and market design experiments:} compare distri-bution-level procurement methods, aggregator access rules, TSO--DSO coordination models, and dynamic tariffs to balance innovation with reliability-first practices.
\end{enumerate}

Millions of flexible devices (EVs, heat pumps, batteries, smart buildings, and industrial processes) will increasingly shape real-time grid conditions. In light of the quantified system potential and demonstrated field results summarized above, \textbf{we therefore issue a call to action:} researchers, system operators, utilities, aggregators, policymakers, and standards bodies across both sides of the Atlantic should jointly build and evaluate interoperable, grid-aware coordination pathways through federated pilots, harmonized interoperability profiles, and coordinated market and policy experiments. By aligning cyber-physical control, data infrastructure, standards, and market mechanisms, the United States and Denmark/EU can co-develop scalable architectures that make distributed flexibility an affordable and reliable system resource.

}

\section*{Acknowledgment}
This work was initiated under the Cross-Atlantic Network for Smart Infrastructure (CANSI) program, funded by the Green International Partnerships (GINP) initiative of the Ministry of Higher Education and Science, Denmark (Ref. No. 2084-000478). The authors would especially like to thank Torben Orla Nielsen of the Foreign Ministry of Denmark for his tireless efforts in leading the CANSI activities, as well as Energy Cluster Denmark and Innovation Centre Denmark, Boston, for their support.

M. Almassalkhi would like to acknowledge that this material is based upon work partially supported by the National Science Foundation under Award No. ECCS-2047306. Any opinions, findings, and conclusions or recommendations expressed in this material are those of the author(s) and do not necessarily reflect the views of the U.S. National Science Foundation.  D. Gayme and Y. Dvorkin would like to acknowledge  partial support from the National Science Foundation under Award No. OISE-2330450. 

Co-authors with DTU gratefully acknowledge funding from the Green International Partnerships (GINP), Ministry of Higher Education and Science, Denmark (Ref. No. 2084-000478); the Center for IT-Intelligent Energy Systems (CITIES), Innovation Fund Denmark (IFD No. 1305-00027B); Flexible Energy Denmark (FED), Innovation Fund Denmark (IFD No. 8090-00069B); and ELEXIA, Horizon Europe (No. 101075656).

\bibliographystyle{elsarticle-num}       
\bibliography{FlexRefs}

\end{document}